\documentstyle[amsmath,epsf,twoside,fleqn,espcrc2]{article}

\newcommand{\tkl}{{{\small\sf T}{$\chi$}{\small\sf L}}}
\newcommand{\sesam}{{\small\sf SESAM}}
\newcommand{\tkll}{{{\sf T}{$\chi$}{\sf L}}}
\newcommand{\sesaml}{{\sf SESAM}}

\newcommand{\bi}{\begin{itemize}}
\newcommand{\ei}{\end{itemize}}
\newcommand{\mpr}{\frac{m_{\pi}}{m_{\rho}}}
\newcommand{\mpv}{\frac{m_{PS}}{m_{V}}}

\newcommand{\beq}{\begin{equation}}
\newcommand{\eeq}{\end{equation}}

\newcommand{\fig}[1]{Fig.~\ref{#1}}
\newcommand{\tab}[1]{Tab.~\ref{#1}}

\newcommand{\eq}[1]{{\frenchspacing Eq.~\ref{#1}}}
\newcommand{\alg}[1]{Algorithm~\ref{#1}}

\newcommand{\oks}{\frac{1}{\kappa_{\rm sea}}}

\newcommand{\okv}{\frac{1}{\kappa_{\rm V}}}
\newcommand{\okvc}{\frac{1}{\kappa_{\rm V}^c}}
\newcommand{\okvl}{\frac{1}{\kappa_{\rm V}^{\rm light}}}
\newcommand{\okc}{\frac{1}{\kappa_{\rm sea}^c}}

\newcommand{\kst}{\kappa^{\rm strange}}
\newcommand{\ks}{\kappa_{\rm sea}}
\newcommand{\kv}{\kappa_{\rm V}}
\newcommand{\kvc}{\kappa_{\rm V}^c}

\newcommand{\ksc}{\kappa_{\rm sea}^c}
\newcommand{\kc}{\kappa_{\rm sea}^c}
\newcommand{\ksl}{\kappa_{\rm sea}^{\rm light}}
\newcommand{\kl}{\kappa_{\rm sea}^{\rm light}}

\newcommand{\mqv}{m^{\rm V}}
\newcommand{\mqs}{m^{\rm sea}}

\newcommand{\mot}{m_{\sf ss}}
\newcommand{\mtf}{m_{\sf vv}}
\newcommand{\moth}{m_{\sf sv}}

\newcommand{\mpsot}{m_{\rm PS, {\sf ss}}^2}
\newcommand{\mpstf}{m_{\rm PS, {\sf vv}}^2}
\newcommand{\mpsoth}{m_{\rm PS, {\sf sv}}^2}

\newcommand{\mvot}{m_{\rm V, {\sf ss}}}
\newcommand{\mvtf}{m_{\rm V, {\sf vv}}}
\newcommand{\mvoth}{m_{\rm V, {\sf sv}}}
\newcommand{\chisq}{\chi^2/{\sf d.o.f}}

\newcommand{\PR}{Phys.\ Rev.~D }
\newcommand{\PRL}{Phys.\ Rev.\ Lett.}

\newcommand{\PL}{Phys.\ Lett.\ }

\newcommand{\vierzehn}{Nucl. Phys. {\bf B} (Proc.\ Suppl.~){\bf 42}
  (1995)}
\newcommand{\prd}[3]{ \frenchspacing Phys.      Rev.  D {\bf #1}
  {(#2)} {#3}} 
\newcommand{\plb}[3]{ \frenchspacing Phys.      Lett. B {\bf #1}
  {(#2)} {#3}}
\newcommand{\ea}{{\frenchspacing et. \frenchspacing al.\hspace{-0.1cm} }}
\newcommand{\eg}{{\frenchspacing\em e.\hspace{0.4mm}g.{}}}
\newcommand{\ie}{{\frenchspacing\em i.\hspace{0.4mm}e.{}}}

\newenvironment{algo}[3]{\vglue6pt\noindent\strut
  \framebox{\begin{minipage}{.45\textwidth}\begin{algorithm} #2 \rm #1\par
        \footnotesize\begin{center}\begin{minipage}{.45\textwidth}
\begin{tabbing}  #3  \end{tabbing}\end{minipage}\end{center}\end{algorithm}\end{minipage}}}{} \newtheorem{algorithm}{Algorithm}

\title{\sesaml\ and \tkll\ Results for Wilson Action--A Status Report}

\author{\frenchspacing Th. Lippert\address{HLRZ, c/o J\"ulich Research
    Center and DESY, Hamburg, D-52425-J\"ulich,
    Germany}\thanks{Presented by Th. Lippert}, G. Bali\address{Physics
    Department, The University, Southampton SO17 1JB, UK}, N.
  Eicker$^{\rm{a}}$, L. Giusti\address{INFN, University ``La
    Sapienza'', P'lle Aldo Moro, Roma, Italy}, U.
  Gl\"assner\address{Department of Physics, University of Wuppertal,
    D-42097 Wuppertal, Germany}, S. G\"usken$^{\rm{d}}$, H.
  Hoeber$^{\rm{d}}$, P.~Lacock$^{\rm{a}}$, G.~Martinelli$^{\rm{c}}$,
  F.  Rapuano$^{\rm{c}}$, G.  Ritzenh\"ofer$^{\rm{a}}$, K.
  Schilling$^{\rm{a,d}}$, G.  Siegert$^{\rm{a}}$, A. Spitz$^{\rm{a}}$,
  P. Ueberholz$^{\rm d}$, and J.  Viehoff$^{\rm{d}}$}
\begin{document}

\begin{abstract}
  Results from two studies of full QCD with two flavours of dynamical
  Wilson fermions are presented. At $\beta=5.6$, the region $0.83 >
  \frac{m_{\pi}}{m_{\rho}} > 0.56$ at $m_{\pi}a > (0.23\mbox L)^{-1}$
  is explored.  The \sesam\ collaboration has generated ensembles of
  about 200 statistically independent configurations on a $16^3\times
  32$-lattice at three different $\kappa$-values and is entering the
  final phase of data analysis.  The \tkl\ simulation on a $24^3\times
  40$-lattice at two $\kappa$-values has reached half statistics and
  data analysis has started recently, hence most results presented
  here are preliminary.  The focus of this report is threefold: ({\em
    i}) we demonstrate that algorithmic improvements like fast Krylov
  solvers and parallel preconditioning recently introduced can be put
  into practise in full QCD simulations, ({\em ii}) we present
  encouraging observations as to the critical dynamics of the Hybrid
  Monte Carlo algorithm in the approach to the chiral limit, ({\em
    iii}) we mention signal improvements of noisy estimator techniques
  for disconnected diagrams to the $\pi$-$N$ $\sigma$ term, and ({\em
    iv}) we report on \sesam's results for light hadron spectrum,
  light quark masses, and heavy quarkonia.
\end{abstract}
\maketitle

\section{INTRODUCTION}

In this talk I will give an interim status report about two
large-scale computer simulations of full lattice QCD with two
degenerate flavours of dynamical Wilson fermions.  Both simulations,
\sesam\ and \tkl, are based on the Hybrid Monte Carlo algorithm
\cite{DUANE,GOTTLIEB} at an inverse coupling of
$\beta=\frac{6}{g^2}=5.6$, running on $16^3\times 32$ and $24^3\times
40$ lattices, respectively.  Our platform is the parallel
supercomputer APE100/Quadrics. The \sesam\ simulation took place on a
256-node 12.8 Gflops machine, while the \tkl\ production is still
ongoing on two 512-node 25.6 Gflops system.  Our goal is to reveal the
effects of dynamical quarks on physical quantities, hence we operate
as close as possible to the chiral limit.  It goes without saying that
this task requires large lattices and high statistics.  Therefore both
{\em technical} and {\em algorithmic} improvements are crucial to
achieve the progress needed. As we Europeans are still living in the
Pre-Teraflops era, we have put much emphasis on the {\em improvement}
of numerical algorithms and the verification of their {\em stochastic
  efficiency}.  I will show in this contribution that we ({\em i})
could boost our simulation speed substantially and ({\em ii}) perform
a reliable determination of the autocorrelation time $\tau$ of the HMC
algorithm.  Another important issue of our work is signal preparation
in the computation of disconnected diagrams to the $\pi$-$N$ $\sigma$
term.  The last part of my talk is devoted to physics results on light
hadron and quark masses and heavy quarkonia.

\subsection{\sesaml}

The name `\sesam' is our slogan and magic acronym in the search for
{\bf S}ea Quark {\bf E}ffects on {\bf S}pectrum {\bf A}nd {\bf M}atrix
Elements in full QCD with dynamical Wilson fermions.  This is indeed a
Tera-computing task.  In order to meet the challenge within the
resources available to us we decided to head for a landmark by aiming
at high statistics at {\em one} value of $\beta$ rather than
attempting a full scaling investigation \cite{GUPTA}.  Needless to say
that, with Pre-Teraflops machines, it is highly non-trivial to
position the hopping parameter window for the observation of sea quark
effects.

The sea-quark masses and lattice resolutions in the \sesam\ simulation
were chosen as small as possible in order to be sensitive to sea-quark
effects and not be disturbed by scaling violations.  Based on
experience from quenched simulations we aimed at lattice spacings of
$a \approx 0.1$ fm corresponding to the quenched $\beta \approx 6.0$
in order to make contact to the scaling region.  The results of the
SCRI-group on the full-QCD $\beta$-function \cite{BITAR} suggested to
go beyond $\beta = 5.5$ in order to escape the strong coupling regime.
To end up with a reasonable physical lattice size---from valence quark
studies at all three different $\ks$'s---we chose to work at $\beta =
5.6$.

\begin{figure}[htb]
\epsfxsize=.465\textwidth\epsfbox{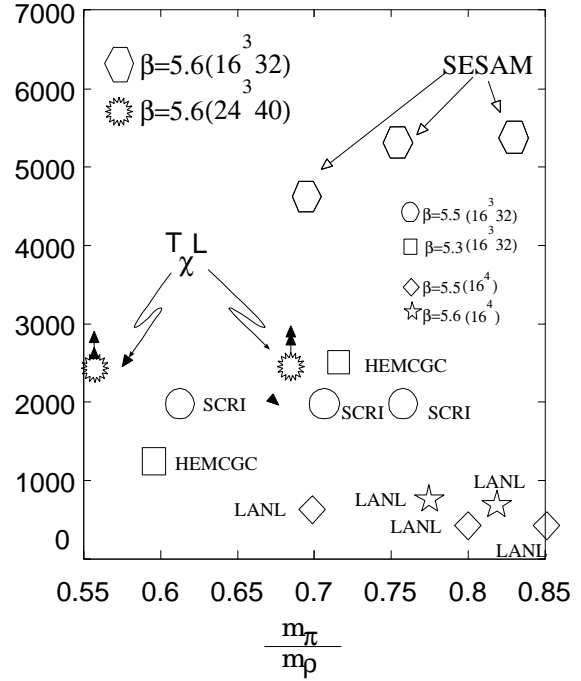}
\caption{\sesam\ and \tkl\ statistics as a function of
  the control parameter $m_\pi/m_\rho$, compared to preceding full QCD
  simulations with Wilson fermions. \label{OVERFIG}}
\end{figure}

In the tuning of $\kappa$ towards its chiral regime, we aim at small
values of $m_\pi/m_\rho$, under the condition that finite-size effects
remain tolerable.  In fact we required $m_{\pi}a L \ge 4$.  This limits
the smallest $\mpr$ ratio that can be attained.  We approached this
minimal $\mpr$ in an iterative manner, starting with our largest bare
mass.  We took into account the statistics that could be achieved
given 1 year runtime on a 256 node Quadrics QH2, extrapolating the
algorithmic specifications from \cite{GUPTA,HEMCGC}.  Eventually we
have generated ensembles at 3 dynamical $\kappa$-values, $\kappa_{sea}
= 0.1560, 0.1570$ and $0.1575$.  Production ended in December 1996.
Altogether, for the \sesam\ project, we have spent on APE slightly
more than 100 Teraflops-hours.  In order to put the statistics
achieved with APE into perspective we present an overview of the
characteristics of recent full QCD simulations with dynamical Wilson
fermions in \tab{OVER} and in \fig{OVERFIG}.
\begin{table*}[tb]
\setlength{\tabcolsep}{1.5pc}
\newlength{\digitwidth} \settowidth{\digitwidth}{\rm 0}
\catcode`?=\active \def?{\kern\digitwidth}
\caption{Some characteristic figures from \sesam, \tkl\ and 
  previously performed large scale simulations with two flavours of
  dynamical Wilson fermions.}
\label{OVER}
\begin{tabular*}{\textwidth}{@{}|l@{\extracolsep{\fill}}|l|l|l|l|r|l|}
\hline
group      & size     &$\beta$ & ${\kappa}$  &$L_S[\mbox{fm}]$ &$L_{traj}$& machine \\\hline\hline
\sesam     & $16^332$ &5.6  & 0.156   &1.44(5) & 5400   & QH2 \\ 
           &          &     & 0.157   &1.39(4) & 5350   &     \\ 
           &          &     & 0.1575  &1.32(4) & 4500   &     \\ 
\hline \hline
\tkl       & $24^340$ &5.6  & 0.1575  &1.93    & 2500   & QH4 \\ 
           &          &     & 0.1580  &        & 2400   &     \\ 
\hline \hline
LANL\cite{GUPTA}       & $16^4$   &5.4  & 0.1600  &        & 564    & CM2 \\ 
           &          &     & 0.1610  & 2.16   & 972    &     \\ 
           &          &     & 0.1620  &        & 239    &     \\ 
\hline
           &          &5.5  & 0.1580  &        & 301            & CM2   \\ 
           &          &     & 0.1590  &1.664   & 396            &           \\ 
           &          &     & 0.1600  &        & 514            &\\ 
\hline
           &          &5.6  & 0.1560  &1.232   & 601            & CM2 \\ 
           &          &     & 0.1570  &        & 756            &     \\ 
\hline
           & $16^332$ &5.5  & 0.1600  &        & 231            & CM2\\ 
\hline
           &          &5.6  & 0.1570  &        & 318            & CM2 \\ 
           &          &     & 0.1575  &        & 168            &\\ 
\hline \hline
HEMCGC\cite{HEMCGC}     & $16^332$ &5.3  & 0.1670  &1.744   & 2425           & CM2 \\ 
           &          &     & 0.1575  &        & 1270           &      \\ 
\hline 
SCRI\cite{SCRI}       & $16^332$ &5.5  & 0.1596  &        &$\sim 2000$     & CM2   \\ 
           &          &     & 0.1600  &        &$\sim 2000$     &           \\ 
           &          &     & 0.1604  &        &$\sim 2000$     &     \\ 
\hline
\end{tabular*}
\end{table*}

The \sesam\ run parameters together with some major monitoring
quantities are given in \tab{RUNSESAM}.  The average number of
iterations is denoted by $\#$ {\em it}, while $N_{md}$ stands for the
number of molecular dynamics steps for a trajectory, which varies
stochastically according to a symmetric distribution of width $\sigma$
in order to avoid interlocks with Fourier modes.
$r=\frac{|Mx-\phi|}{|x|}$ is the stopping criterion for the iterative
solver, {\em alg\,} specifies the inversion algorithm used, see
sections \ref{INVERTER}. and \ref{PRECONDITIONING}.
\begin{table*}[t]
\setlength{\tabcolsep}{.3pc}
\caption{Simulation parameters and some monitor quantities for the
  \sesam\ runs. }
\label{RUNSESAM}
\begin{tabular*}{\textwidth}{@{}|l@{\extracolsep{\fill}}|l|l|l|l|l|l|l|l|l|l|}
\hline 
$\kappa$   &  alg  &  $T$   &  $N_{md} \pm \sigma_{N_{md}}$    
 & acc [\%] &  r  & \# it  & $N_{CSG}$ & $\mpr$   &  \# traj & \# therm \\ 
\hline\hline 
0.156                    & o/e  & 1    & 100 $\pm$ 20 & 85 &
$10^{-8}$ & 85(3)    &  6  &                  {0.8388(41)} & 5000  &  400 \\
\hline
\raisebox{-.3cm}{0.1570} & o/e  & 1    & 100 $\pm$ 20 & 84 & 
$10^{-8}$ & 168(5)   &  8  & \raisebox{-.3cm}{0.7552(69)}   & 3500  &  350  \\
                         & SSOR & 1    & 100 $\pm$ 20 & 80 &
$10^{-8}$ & 125(3)   &  9  &                          & 1500  &  0 \\
\hline 
\raisebox{-.3cm}{0.1575} & o/e  & 1    & 100 $\pm$ 20 & 76 & 
$10^{-8}$ & 317(12)  &  11 &  \raisebox{-.3cm}{0.688(12)} &  3000  & 500   \\
                         & SSOR & 0.5  & 71  $\pm$ 12 & 73 &
$10^{-8}$ & 150(6)   &  3  &                          &  2000  & 0\\
\hline
\end{tabular*} 
\end{table*}
After thermalization of each $\kappa$-run we have held the molecular
dynamics parameters fixed; this enables us to carry out a sensible
autocorrelation analysis. We found that switching
$\kappa$, while holding the molecular dynamics time step fixed, is
affecting the acceptance rate only marginally, see \tab{RUNSESAM}.
Therefore we have used a universal time step of $dt=0.01$ for most
dynamical samples.

It was one of the goals of \sesam\ to perform for the first time a
detailed autocorrelation study on an adequate set of physical
quantities.  To this end we decided to archive all HMC trajectories
and study more difficult observables, such as glueballs and
topological features, in postprocessing style using the Crays T3E, T90
and Personal Computers at HLRZ J\"ulich.

Simulations of full QCD with two flavours of Wilson fermions at zero
temperature so far have been carried out on lattices of size $\le
16^{3}\times 32$ and ratios of $\mpr > 0.6$
\cite{GUPTA,HEMCGC,SCRI,SESAM_MEL}.  The results of \sesam\ 
demonstrate that dedicated supercomputers in the range of about 10
Gflops performance can indeed generate in one year's runtime
statistically significant full QCD samples at $\mpv\simeq 0.70$. Alas,
according to $\chi$PT (constructing a fictitious
pseudoscalar meson containing two `strange quarks', with mass ratio of
the size quoted),
\begin{equation} 
\frac{m_{ps}}{m_{\phi}}\approx
\frac{\sqrt{2m^{2}_{K}}}{m_{\phi}}=0.69,
\label{CHI}
\end{equation}
we find ourselves still in the region of strange quarks. Thus, in
order to quantify light sea quark effects in full QCD, one would wish
to come closer to the chiral limit and to finer lattice resolutions than
achieved previously. This implies larger lattice volumes.

\subsection{\tkll}

In a feasibility study on a $24^3 \times 40$ lattice we have
investigated whether a further step towards the chiral limit is in
reach of the APE Tower computing power.  We tuned ourselves to a
realistic working point at a volume of (2 fm)$^{3}$, with chirality
characterised by $\frac{1}{{m}_{\pi}a} \approx 5.6$ and $\mpr<0.6$
\cite{LIPPERT_LOUIS}.  We found that by use of preconditioning
techniques we could accelerate the matrix inversion
\cite{FISCHER,GERO} sufficiently for a 512-node APE Tower to drive an
optimised HMC code fast enough ({\em i}) to increase the lattice size
by more than a factor of 4 compared to the previous standards
including \sesam, ({\em ii}) and to go more chiral.

In the framework of the Italian-German \tkl-collaboration we launched
an 18 month Hybrid Monte Carlo simulation, mainly running on the APE100
Tower at INFN Rome and partly on the QH4 Zeuthen/Berlin at
$\beta=5.6$.  Our HMC implementation reaches a sustained performance
of 17 Gflops with 25.6 Gflops being the APE Tower peak speed.

We switched from the thinned o/e representation to the full representation
of the fermionic matrix $M={\bf 1}-\kappa D$, employing our new SSOR
preconditioning scheme, see section \ref{PRECONDITIONING}. On the APE
machine, this scheme offers about an overall gain of a factor of 2 in
machine time, see \tab{RUNTKL}.

We have chosen two $\kappa$-values, 0.1575 and 0.158 for two reasons:
({\em i}) $\kappa=0.1575$ coincides with the smallest mass value of
the \sesam\ project relating the two lattice sizes at a definite point
in parameter space.  Thus we are able to assess both the influence of
finite size effects on physical quantities and the volume dependence
of the simulation algorithm. ({\em ii}) The $24^3\times 40$ lattice
allows to increase the pion correlation length in lattice units,
$\xi_{\pi}$, by a factor of 1.5 compared to the smallest dynamical
mass of the \sesam\ project, which we estimated to be sufficient for a
chirality $\mpr$ in the range of $.55$.  As the lattice scale is
refined when reducing the bare quark masses we decided to stay with
$\beta=5.6$.

\begin{figure}[htb]
\epsfxsize=.465\textwidth\epsfbox{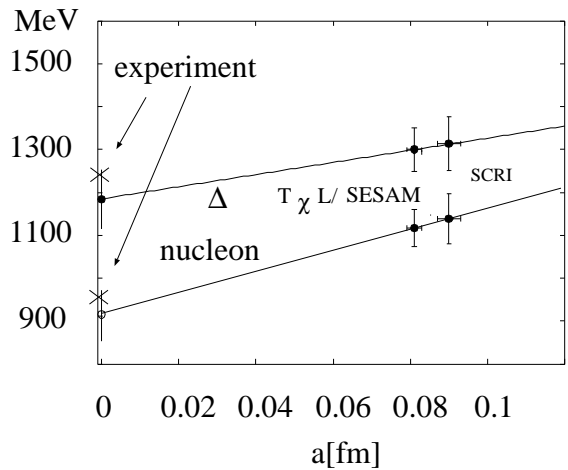}
\caption{
  ``Continuum extrapolation'' for baryon masses (together with SCRI
  data \cite{SCRI}) obtained at $\beta=5.5$.
\label{SCALING}}
\end{figure}

In order to perform a scaling analysis it would have been desirable
to simulate at different $\beta$-values.  The peephole to scaling,
{\em cf.\ } \fig{SCALING}, discloses unfortunately that a substantial
decrease in scale $a$ would be needed in order to have a sufficient
lever arm in the continuum extrapolation.  It is instructive though to
compare \fig{SCALING} to recent improved action results
\cite{KENWAY_thisreport}.

\begin{figure}[htb]
\epsfxsize=.465\textwidth\epsfbox{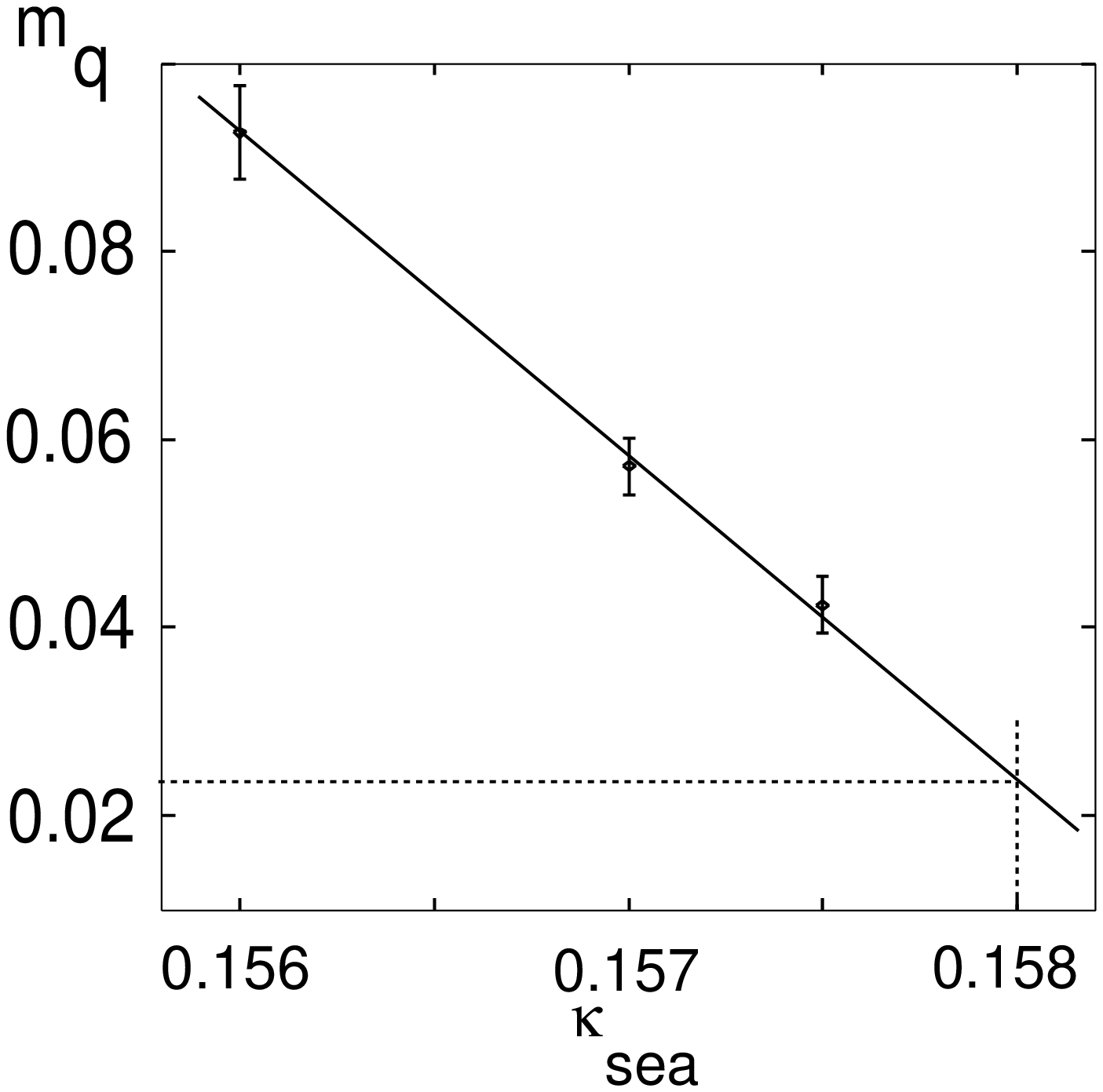}
\caption{
  Fixing $\ks$ for \tkl\ from \sesam\ data.
\label{EXTRA}}
\end{figure}

During the layout phase of \tkl\ (Feb.\ 1996) we determined the most
chiral $\ks$ value extrapolating the relation $m_qa=\frac{1}{2}\Big(
\frac{1}{\kappa}- \frac{1}{\kc} \Big)$ on the \sesam\ data available
at the time.  Requiring $\frac{\xi_{\pi}}{24a} = .23$ we arrived at
$m_q\,a=0.023$ corresponding to $\kappa=0.1580$, {\em cf.}\ 
\fig{EXTRA}.  We considered this value of $\xi_{\pi}$ as small enough
not to suffer from large finite size effects.
Needless to say: at the end of the day it remains to be seen, see
section \ref{SPECTRUM}, that this parameter choice is reasonably
positioned within the `chirality gap'. At the time of this meeting, we
have generated two ensembles of configurations at $\kappa=0.1575$ and
$\kappa=0.158$, with more than 2400 trajectories each, within 150
Teraflops-hours.  For \tkl\ we were not able to archive each
trajectory so we decided to store every other for later
autocorrelation analysis and detailed postprocessing.

The \tkl\ run parameters and some monitoring quantities of interest are
collected in \tab{RUNTKL}.  $t$ specifies the time to generate one
trajectory on the APE Tower.
\begin{table*}[t]
\setlength{\tabcolsep}{.3pc}
\caption{Simulation parameters and some monitor  quantities for the
  \tkl\ runs. }
\label{RUNTKL}
\begin{tabular*}{\textwidth}{@{}|l@{\extracolsep{\fill}}|l|l|l|l|l|l|l|l|}
  \hline $\kappa$ & alg & $T$ & $N_{md} \pm \sigma_{N_{md}}$ & r &
  t/traj [s] & $\mpr$ [prel!]  & \# traj & \# therm \\ \hline\hline
  \raisebox{-.3cm}{0.1575} & o/e & 1 & 125 $\pm$ 20 & $10^{-8}$ & 8200 & &
  & \\ & SSOR & 0.5 & 125 $\pm$ 20 & $10^{-8}$ & 3800 & 0.70(4) & 2500 &
  500 \\ \hline 0.158 & SSOR & 0.5 & 125 $\pm$ 20 & $10^{-8}$ & 9200 &
  0.56(4) & 2400 & 800 \\ \hline
\end{tabular*} 
\end{table*}

\section{OPTIMIZING HMC}

The optimisation of fermionic simulation algorithms constitutes a
major target of the scientific program of the Wuppertal-HLRZ lattice
QCD group.  From the beginning the two projects, \sesam\ and \tkl,
have been accompanied by algorithmic research in collaboration with
the Applied Mathematics group at Wuppertal University. We have focused
on the acceleration of Krylov subspace solvers within the computer
intensive inversion part of HMC that is required to calculate the
fermionic force \cite{FROMMER1,FROMMER2}. In these studies the
so-called biconjugate gradient stabilised method (BiCGstab) has been
established as the most efficient Krylov solver for Wilson fermion
matrix inversions. It behaves quasi-optimal, {\em i.e.}, it approaches
the convergence speed of GMRES, the (non-practical) optimal reference
Krylov solver.  Therefore, to make further progress, one has to
address the development of new parallel preconditioning methods
\cite{FISCHER,FROMMER_LOUIS}.

\subsection{Krylov solvers\label{INVERTER}}

The practical iterative methods to solve the huge system of equations
$MX=\phi$ belong to the class of Krylov subspace methods. In the old
days the minimum residual algorithm (MR) has been established as {\em
  the} efficient method for both propagator computations and the
calculation of the fermionic force within the HMC \cite{GUPTAOLD}. In
the early nineties, numerical analysts were successful in developing
new Krylov subspace methods \cite{QMR,VORST}.  These methods avoid the
squared condition number (from $M^{\dagger}M$ inversion) of CG, and
yet guarantee convergence as opposed to MR.

In Refs.~\cite{FROMMER1,FROMMER2,GLAESSNER1} we have benchmarked
and improved various iterative solvers (within simplified settings) in
order to find the fastest solver for Wilson fermion matrix inversions.

\begin{figure}[htb]
\label{CONVERGENCE}
\epsfxsize=.465\textwidth\epsfbox{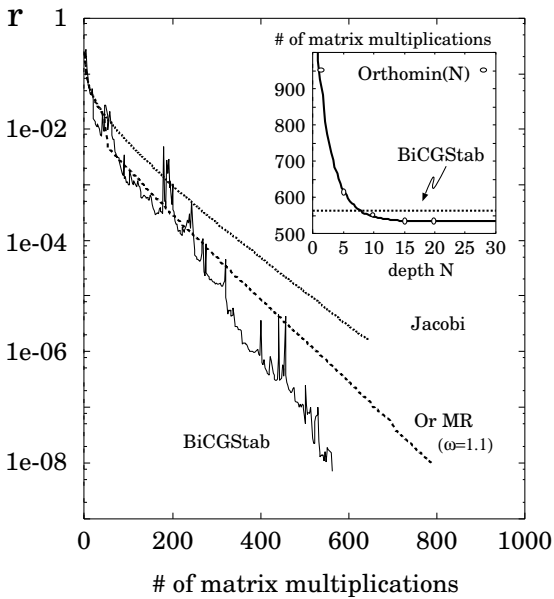}
\caption{Comparing the convergence of 
  the simple Jacobi iteration, ORMR and BiCGstab at $\beta=5.6$ and
  $\kappa=0.1575$ for a typical $16^{3}\times 32$ full QCD configuration.
  The insertion shows the results for OrthoMin(N) compared to BiCGstab
  (straight line).}
\end{figure}

Using \sesam's configurations we can confirm the findings of
\cite{FROMMER1}.  The over-relaxed MR algorithm outperforms CG, however
BiCGstab beats ORMR, see \fig{CONVERGENCE}. This behaviour is
qualitatively the same for all three $\ks$-values used by \sesam.

Comparing the convergence behaviour of BiCGstab with that of
OrthoMin(N) reveals that BiCGstab is close to the optimal algorithm
GMRES. GMRES orthonormalises on all previous search directions and
therefore is not practical. OrthoMin(N) \cite{ORTHO} is a practical
modification of the latter, orthogonalising on $N$ previous directions
only. The insertion demonstrates that BiCGstab is beating OrthoMin up
to a depth of $N=10$ and further on is about 10 \% less efficient in
terms of iterations. In view of these findings, BiCGstab can be
considered as {\em quasi optimal}\footnote{In terms of computer time
  expense, BiCHstab is more efficient.}.

\subsection{Parallel preconditioning\label{PRECONDITIONING}}

The quasi-optimality of BiCGstab suggests to turn attention on
multigrid methods and/or preconditioning in order to achieve further
speed up in Wilson fermion matrix inversions.  The application of
multigrid techniques is impractical, however, due to the gauge noise
of the gluonic background field entering the fermion matrix.  Thus
preconditioning techniques, \ie\ methods to decrease the condition
number $K^2$ of $M^{\dagger}M$ appear to be the only promising path to
further accelerate Krylov subspace solvers like BiCGstab.

A standard preconditioning approach in lattice gauge computations rests
upon o/e decomposition of $M$ \cite{ROSSI}.  It can yield an efficiency
gain by a factor of 2 when inverting $M$.  Some years ago, Oyanagi
\cite{Oy85} exploited incomplete LU (ILU) factorisation of the matrix $M$
based on the natural, {\em globally lexicographic}\/ ordering of the
lattice points\footnote{For the Wilson fermion discretisation with Wilson
  parameter $r=1$, ILU preconditioning is identical to symmetric successive
  over-relaxation (SSOR) preconditioning with respect to that ordering.}.
On local memory or grid-oriented parallel computers, this preconditioner
can hardly be implemented efficiently, however.

Recently we have introduced a new parallel preconditioning technique
suitable for Wilson fermion matrix inversions.  Our method is called {\em
  local lexicographic SSOR preconditioner} (LL-SSOR).  As opposed to
familiar multicolour SSOR preconditioners (like the o/e preconditioner)
which produce a decoupling of variables on a very fine grain level, the
LL-SSOR method provides the flexibility to reduce the decoupling to the
minimum which is necessary to suit a given parallel system.  As for any
SSOR preconditioner, the Eisenstat Trick \cite{EISENSTAT} is crucial for the
efficient implementation of LL-SSOR.  Our numerical experiences show that
LL-SSOR presently offers the fastest available solution method on parallel
computers.

The general preconditioning of $M$ proceeds via two non-singular
matrices $V_1$ and $V_2$, the left and right preconditioners,
respectively:
\begin{eqnarray}
V_1^{-1}MV_2^{-1} \tilde{x} &=& \tilde{\phi},\nonumber\\
\mbox{where } 
\tilde{\phi} &=& V_1^{-1}\phi,
\; \tilde{x} = V_2x.
\end{eqnarray}
A Krylov solver could now be applied directly, by replacement of each
occurrence of $M$ and $\phi$ by $V_1^{-1}MV_2^{-1}$ and $\tilde{\phi}$.

We have chosen to apply symmetric Gau\ss{}-Seidel (SSOR) preconditioning.
The matrix $M$ has to be decomposed into its diagonal, strictly lower and
strictly upper triangular parts, $ M = I - L - U$. Then the SSOR
preconditioner is specified by
\begin{equation}
\label{SSOR_prec_eq}
V_1 = I - L, \enspace V_2 = I - U.
\end{equation}

Now we find that $V_1 + V_2 - M = I$. This relation allows to exploit the
Eisenstat-trick \cite{EISENSTAT}:  $V_1^{-1}MV_2^{-1} = V_2^{-1} +
V_1^{-1}(I-V_2^{-1})$, so that the matrix vector product $w =
V_1^{-1}MV_2^{-1}r$ amounts to a 2-step computation
\begin{equation}
v = V_2^{-1}r, \enspace u = V_1^{-1}(r-v), \enspace w = v + u.
\end{equation}
The general preconditioned BiCGstab procedure is described in
Ref.~\cite{FISCHER}.

Since the matrices $I-L$ and $I-U$ are triangular, their inversions are
simply performed by {\em forward} and {\em backward} substitution,
respectively.  As example the forward solve $(I- L)y = p$ becomes:
\begin{tabbing}
\hspace*{2ex} \= \hspace{2ex} \= \kill
for $i=1,\ldots,n$ \\
\> $\displaystyle (y)_i = (p)_i + \sum_{j=1}^{i-1} (L)_{ij}(y)_j$.
\end{tabbing}
In terms of computational cost, a forward followed by a backward solve is
as expensive as a multiplication with $M$.  Hence in principle there is no
increase of computational cost with SSOR.

In the solution of the Wilson fermion inversion problem the ordering scheme
for the lattice points $x$ is completely  arbitrary. Different orderings yield
different matrices $M$, permutationally similar to each other.  The
efficiency of the SSOR preconditioner depends on the ordering scheme
chosen.

Consider an arbitrary numbering (ordering) of the lattice points. For a
given grid point $x$, the corresponding row in the matrix $L$ or $U$
contains exactly the coupling coefficients of those nearest neighbours of
$x$ which have been numbered before or after $x$, resp.  Therefore, a
generic formulation of the forward solve for this ordering is given by
Algorithm~\ref{generic-forward}.  The backward solves are done similarly,
now running through the grid points in {\em reverse} order.
\begin{algo}{Forward Solve.}{\label{generic-forward}}{
\hspace*{2ex} \= \hspace{2ex} \=  \hspace{2ex} \= \hspace{2ex} \kill
for all grid points $x$ in the given order \\
\> \{ update $y_x$ \} \\
\> $y_x = p_x $ \\
\> for $\mu = 1,\ldots,4$ \\
\>  \> if $x-\mu$ was numbered before $x$ then \\
\>  \>  \> $\displaystyle y_x = y_x + \kappa \cdot  m^+_{x,x-\mu}y_{x - \mu}$ \\
\> for $\mu = 1,\ldots,4$ \\
\>  \> if $x+\mu$ was numbered before $x$ then \\
\>  \>  \> $\displaystyle y_x = y_x + \kappa \cdot  m^-_{x,x+\mu}y_{x + \mu}$ 
}
\end{algo}

So far, odd-even preconditioning was considered as the only successful
preconditioner for lattice QCD that is parallelisable.  For this
particular ordering the inverses of $I-L$ and $I-U$ can be determined
directly, see Ref.~\cite{FISCHER}.

Oyanagi some time ago \cite{Oy85} has demonstrated that ILU (SSOR)
preconditioning, applying global lexicographic ordering, yields an
improvement over odd-even preconditioning as far as the number of
iterations is concerned.  However, its parallel implementation is difficult
\cite{HOCKNEY}.

The ordering we have proposed is similar to Oyanagi's approach,
however, it is intrinsically parallelisable, and it is adaptive
to the parallel computer used. We assume that the processors of the
parallel computer are connected as a $p_1 \times p_2 \times p_3 \times
p_4$ 4-dimensional grid.  The space-time lattice can be matched to the
processor grid in an obvious natural manner, producing a local lattice
of size $n^{loc}_1 \times n^{loc}_2 \times n^{loc}_3 \times n^{loc}_4$
with $n^{loc}_i = n_i/p_i$ on each processor.

\begin{figure}[htb]
\setlength{\unitlength}{0.6cm}
\begin{center}
\begin{picture}(12,10)
\multiput(0,0)(0,1){10}{\multiput(0,0)(1,0){12}{\circle{0.5}}}
\multiput(1.5,-0.5)(4,0){3}{\line(0,1){10}}
\multiput(-0.5,2.5)(0,4){2}{\line(1,0){12}}
\put(2.25,6){\vector(1,0){0.5}}
\put(2,4.75){\vector(0,-1){0.5}}
\put(4.25,5){\vector(1,0){0.5}}
\put(5,5.75){\vector(0,-1){0.5}}
\put(5.75,5){\vector(-1,0){0.5}}
\put(8,4.75){\vector(0,-1){0.5}}
\put(7.25,4){\vector(1,0){0.5}}
\put(4.25,3){\vector(1,0){0.5}}
\put(5,2.25){\vector(0,1){0.5}}
\put(5,3.75){\vector(0,-1){0.5}}
\put(5.75,3){\vector(-1,0){0.5}}
\put(6.25,4){\vector(1,0){0.5}}
\put(7,4.75){\vector(0,-1){0.5}}
\begin{tiny}
\begin{em}
\put(0,9){\makebox(0,0){g}}
\put(1,9){\makebox(0,0){h}}
\put(2,9){\makebox(0,0){e}}
\put(3,9){\makebox(0,0){f}}
\put(4,9){\makebox(0,0){g}}
\put(5,9){\makebox(0,0){h}}
\put(6,9){\makebox(0,0){e}}
\put(7,9){\makebox(0,0){f}}
\put(8,9){\makebox(0,0){g}}
\put(9,9){\makebox(0,0){h}}
\put(10,9){\makebox(0,0){e}}
\put(11,9){\makebox(0,0){f}}
\put(0,8){\makebox(0,0){l}}
\put(1,8){\makebox(0,0){m}}
\put(2,8){\makebox(0,0){i}}
\put(3,8){\makebox(0,0){k}}
\put(4,8){\makebox(0,0){l}}
\put(5,8){\makebox(0,0){m}}
\put(6,8){\makebox(0,0){i}}
\put(7,8){\makebox(0,0){k}}
\put(8,8){\makebox(0,0){l}}
\put(9,8){\makebox(0,0){m}}
\put(10,8){\makebox(0,0){i}}
\put(11,8){\makebox(0,0){k}}
\put(0,7){\makebox(0,0){p}}
\put(1,7){\makebox(0,0){q}}
\put(2,7){\makebox(0,0){n}}
\put(3,7){\makebox(0,0){o}}
\put(4,7){\makebox(0,0){p}}
\put(5,7){\makebox(0,0){q}}
\put(6,7){\makebox(0,0){n}}
\put(7,7){\makebox(0,0){o}}
\put(8,7){\makebox(0,0){p}}
\put(9,7){\makebox(0,0){q}}
\put(10,7){\makebox(0,0){n}}
\put(11,7){\makebox(0,0){o}}
\put(0,6){\makebox(0,0){c}}
\put(1,6){\makebox(0,0){d}}
\put(2,6){\makebox(0,0){a}}
\put(3,6){\makebox(0,0){b}}
\put(4,6){\makebox(0,0){c}}
\put(5,6){\makebox(0,0){d}}
\put(6,6){\makebox(0,0){a}}
\put(7,6){\makebox(0,0){b}}
\put(8,6){\makebox(0,0){c}}
\put(9,6){\makebox(0,0){d}}
\put(10,6){\makebox(0,0){a}}
\put(11,6){\makebox(0,0){b}}
\put(0,5){\makebox(0,0){g}}
\put(1,5){\makebox(0,0){h}}
\put(2,5){\makebox(0,0){e}}
\put(3,5){\makebox(0,0){f}}
\put(4,5){\makebox(0,0){g}}
\put(5,5){\makebox(0,0){h}}
\put(6,5){\makebox(0,0){e}}
\put(7,5){\makebox(0,0){f}}
\put(8,5){\makebox(0,0){g}}
\put(9,5){\makebox(0,0){h}}
\put(10,5){\makebox(0,0){e}}
\put(11,5){\makebox(0,0){f}}
\put(0,4){\makebox(0,0){l}}
\put(1,4){\makebox(0,0){m}}
\put(2,4){\makebox(0,0){i}}
\put(3,4){\makebox(0,0){k}}
\put(4,4){\makebox(0,0){l}}
\put(5,4){\makebox(0,0){m}}
\put(6,4){\makebox(0,0){i}}
\put(7,4){\makebox(0,0){k}}
\put(8,4){\makebox(0,0){l}}
\put(9,4){\makebox(0,0){m}}
\put(10,4){\makebox(0,0){i}}
\put(11,4){\makebox(0,0){k}}
\put(0,3){\makebox(0,0){p}}
\put(1,3){\makebox(0,0){q}}
\put(2,3){\makebox(0,0){n}}
\put(3,3){\makebox(0,0){o}}
\put(4,3){\makebox(0,0){p}}
\put(5,3){\makebox(0,0){q}}
\put(6,3){\makebox(0,0){n}}
\put(7,3){\makebox(0,0){o}}
\put(8,3){\makebox(0,0){p}}
\put(9,3){\makebox(0,0){q}}
\put(10,3){\makebox(0,0){n}}
\put(11,3){\makebox(0,0){o}}
\put(0,2){\makebox(0,0){c}}
\put(1,2){\makebox(0,0){d}}
\put(2,2){\makebox(0,0){a}}
\put(3,2){\makebox(0,0){b}}
\put(4,2){\makebox(0,0){c}}
\put(5,2){\makebox(0,0){d}}
\put(6,2){\makebox(0,0){a}}
\put(7,2){\makebox(0,0){b}}
\put(8,2){\makebox(0,0){c}}
\put(9,2){\makebox(0,0){d}}
\put(10,2){\makebox(0,0){a}}
\put(11,2){\makebox(0,0){b}}
\put(0,1){\makebox(0,0){g}}
\put(1,1){\makebox(0,0){h}}
\put(2,1){\makebox(0,0){e}}
\put(3,1){\makebox(0,0){f}}
\put(4,1){\makebox(0,0){g}}
\put(5,1){\makebox(0,0){h}}
\put(6,1){\makebox(0,0){e}}
\put(7,1){\makebox(0,0){f}}
\put(8,1){\makebox(0,0){g}}
\put(9,1){\makebox(0,0){h}}
\put(10,1){\makebox(0,0){e}}
\put(11,1){\makebox(0,0){f}}
\put(0,0){\makebox(0,0){l}}
\put(1,0){\makebox(0,0){m}}
\put(2,0){\makebox(0,0){i}}
\put(3,0){\makebox(0,0){k}}
\put(4,0){\makebox(0,0){l}}
\put(5,0){\makebox(0,0){m}}
\put(6,0){\makebox(0,0){i}}
\put(7,0){\makebox(0,0){k}}
\put(8,0){\makebox(0,0){l}}
\put(9,0){\makebox(0,0){m}}
\put(10,0){\makebox(0,0){i}}
\put(11,0){\makebox(0,0){k}}
\end{em}
\end{tiny}
\end{picture}
\end{center}
\caption{Locally lexicographic ordering and forward solve in 2 dimensions.
\label{COLORING}} 
\end{figure}
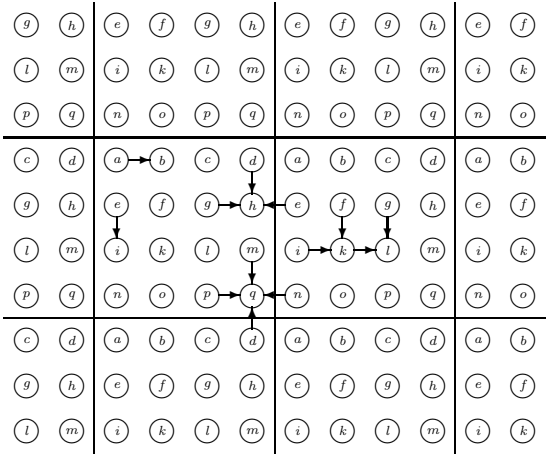

The whole lattice is divided into $n^{loc} = n^{loc}_1 n^{loc}_2
n^{loc}_3 n^{loc}_4$ groups.  Each group corresponds to a fixed
position of a site in the local grid.  Associating a colour with each
of the groups, we can interpret this process as a colouring of the
lattice points, see \fig{COLORING}.

All nearest neighbours of a given grid point have colours different from that
point. Performing the forward and backward solves within the BiCGstab
algorithm, grid points having the same colour can be worked upon in
parallel, thus yielding an optimal parallelism of $p$, the number of
processors.

A formulation of the $ll$-forward solve is given in
\alg{ll-forward}.  Here, we use `$\leq_{ll}$' as a symbol
for `$ll$-less than'.

On the `local boundaries' we will have between 0 (for the $ll$-first point)
and 8 (for the $ll$-last point) summands to add to $p_x$.  The parallelism
achieved is $p$, and thus is optimal since we have $p$ processors.  If we
change the number of processors, the $ll$-ordering, and consequently the
properties of the corresponding SSOR preconditioner will change, too.

\begin{figure}[htb]
\epsfxsize=.465\textwidth\epsfbox{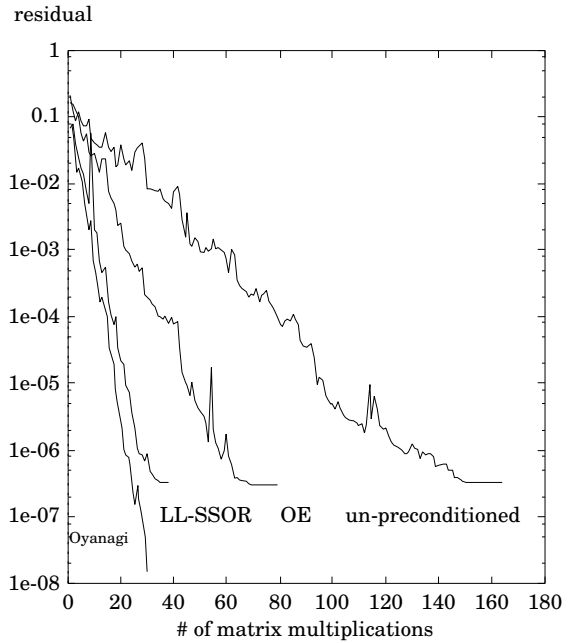}
\caption[]{
Residue reached by a given number of matrix-multiplies.}
\label{RESIDUE} 
\end{figure}

The efficiency of LL-SSOR has been tested in the framework of \sesam\ 
and \tkl\ simulations.  First we display performance results for
$\kappa=0.157$ on an $8^3\times 16$ lattice at $\beta=5.6$.  In
\fig{RESIDUE} we show that the convergence speed of LL-SSOR is about
twice as fast as that of o/e preconditioning and nearly reaches that
of Oyanagi preconditioning. On Quadrics we have achieved a {\em real
  overall speed up} of a factor between 1.5 and 2.1 compared to our o/e
implementation, see \tab{RUNTKL}; however, this is a machine dependent
result.
\begin{algo}{ll-forward.}{\label{ll-forward}}{
\hspace*{2ex} \= \hspace{2ex} \=  \hspace{1.2cm} \= \hspace{2ex} 
\= \hspace{1cm} \kill
for all colours in lexicographic order \\
\> for all processors \\
\>  \> $x := $ grid point of that colour on that processor \\
\>  \> \{ update $y_x$ \} \\
\>  \>  $\displaystyle y_x = p_x + \kappa$ \>\> $\displaystyle \left( 
         \sum_{\mu, \, x-\mu \, \leq_{ll} \, x} m^+_{x,x-\mu}y_{x - \mu}
         \right.$\\
\> \> \>  \>  $ \displaystyle \left.  +   
     \sum_{\mu, \, x+\mu \, \leq_{ll} \, x} m^-_{x,x+\mu}y_{x+\mu}
              \right)$}
\end{algo}

\begin{figure}[htb]
\epsfxsize=.465\textwidth\epsfbox{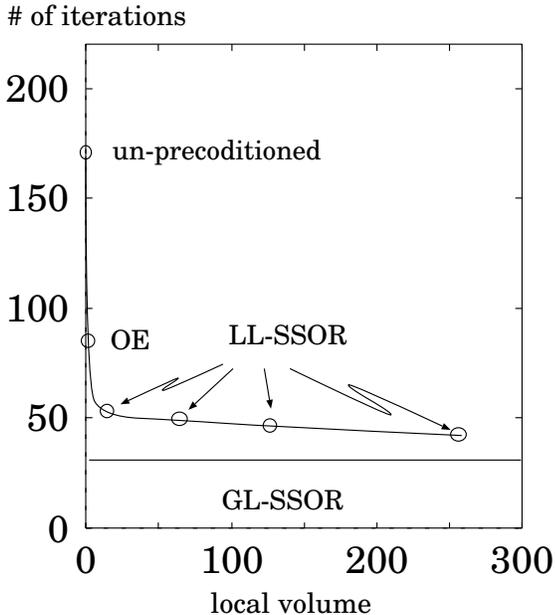}
\caption[]{
Volume dependency of LL-SSOR efficiency for BiCGstab.}
\label{VDEP} 
\end{figure}

We also can present results for the volume dependency of the LL-SSOR
preconditioning, see \fig{VDEP}.

Finally we remark that preconditioning and chronological start vector
guess (CSG) \cite{BROWER} as applied in the molecular dynamics
evolution part of HMC are nearly additive in their iteration gain.

\section{CRITICAL DYNAMICS}

The HMC algorithm is a Markov process at heart. Therefore, a full QCD
vacuum configuration series exhibits autocorrelation. Its statistical
quality is affected crucially by this autocorrelation which depends in
general on the physical observable under investigation and are
increasing in the approach to the chiral limit.

The exact determination of the autocorrelation function amounts to trace
the system over infinite time.  Any practical Markov process is finite,
however, and it is so much the worse that $t_{MC}$ in full QCD Hybrid Monte
Carlo simulations is not large enough compared to the relaxation time,
$t_{MC} \simeq \tau_{exp}$.  So far this has prevented a reliable
determination of autocorrelations in full QCD simulations with Wilson
fermions.

\sesam\ has increased the trajectory samples by nearly one order of
magnitude compared to previous studies\footnote{\tkl\ aims at $O(4000)$
  trajectories per dynamical sample.}.  It is important to note that we
rested under stable conditions for the HMC dynamics to evolve rather than
retuning MD parameters as production went on.  This provides the setting
for a reliable determination of autocorrelation times related to various
gluonic and hadronic quantities.

Given a time-series of measurements $A_t$, $t=1,\dots,t_{MC}$ we compute a
finite time-series approximation to the true autocorrelation function for
observable $A$:
\begin{eqnarray}
C^A(t)= 
\frac{1}{t_{MC}-t}
\sum_{s=1}^{t_{MC}-t} A_sA_{s+t}\nonumber\\
 - \left(
\frac{1}{t_{MC}-t}
\sum_{s=1}^{t_{MC}}
A_s \right) ^2.
\end{eqnarray}
This two-point function in the Monte Carlo time can be normalised by
\begin{equation}
\rho^A(t) = \frac{C^A(t)}{C^A(0)}.
\end{equation}

The {\em exponential} autocorrelation time is defined as the inverse
decay rate of the slowest mode contributing to the autocorrelation
function:
\begin{equation}
\tau^A_{exp} = \limsup_{t\rightarrow\infty}\frac{t}{-\log\rho^A(t)}. 
\end{equation}
$\tau^A_{exp}$ is a relaxation parameter and is related to the length of
the thermalization phase \cite{SOKAL} of the Markov process. In our
simulations we required the thermalization length to be $5 \times
\tau_{exp}$ such as to achieve a suppression of the starting conditions of
order ${\cal O}(\exp(-5))$ in the ensemble.  Furthermore $\tau^A_{exp}$ is
a characteristic time to achieve ergodicity: the simulation has to be much
larger than $\mbox{sup}_A \{\tau^A_{exp}\}$.

The {\em integrated} autocorrelation time is defined as the integral:
\begin{equation}
\tau^A_{int}=\frac{1}{2}+\sum_{t'=1}^{t_{MC} \to \infty}\rho^A(t').
\label{tauint}
\end{equation}
In equilibrium $\tau^A_{int}$ characterises the true statistical
error of the observable $A$ computed from the ensemble.  The sample
mean
\begin{equation}
\langle A\rangle = \frac{1}{N}\sum_{i=1}^N A(i),
\end{equation}
has the variance
\begin{eqnarray}
\sigma_A^2 \approx \frac{1}{N-1}2\tau^A_{int}\,C^A(0)  
= 2\tau^A_{int}\sigma^2_{0},\nonumber\\
\mbox{for $N\gg\tau^A_{int}$},
\end{eqnarray}
which is increased by the factor $2\tau^A_{int}$ compared to the
result over a sample of $N$ independent configurations.  In our
simulations, configurations separated by $2$ to $3\times \tau^A_{int}$
are considered as `decorrelated'\footnote{Residual autocorrelation
  between successive measurements are taken into account by binning
  the data prior to statistical analysis \cite{SPECTRUMPAPER}.}.

On a finite time-series it is difficult to estimate the slowest exponential
autocorrelation time reliably as the tail of the autocorrelation function
becomes compatible with zero. This leads to an unwanted bias in
$\tau^A_{int}$.  A practical solution to this problem is the application of
a `window' procedure as introduced in Ref.~\cite{SOKAL} to extract the
integrated autocorrelation time: a cut-off $t$ in the sum for
$\tau^A_{int}$ is increased until a plateau becomes visible.  As a rule of
thumb, it has been suggested in \cite{SOKAL} to determine $t$
self-consistently in the range $4$ to $10 \tau^A_{int}$.  This amounts to a
truncation effect (difference to the true $\tau_{int}$) of less than $2\%$.

The integrated autocorrelation time $\tau_{int}$ can be observable
dependent. One effect is due to the time-space extension of lattice
observables. Very extended quantities on the lattice might exhibit
larger $\tau_{int}$.  It can be shown in free field models that
autocorrelation modes are related to lattice symmetries like \eg\ 
translation invariance, and in this way to correlations on the
lattice itself.  As a consequence long range correlations across the
lattice, as they occur for light masses, go along with larger
autocorrelation times.

In the following we determine the autocorrelation from a variety of 
observables.  This is facilitated because we have archived all
trajectories of the \sesam-simulation and every second trajectory of
the \tkl-simulation for a detailed post-processing.

\begin{figure}[htb]
\epsfxsize=.465\textwidth\epsfbox{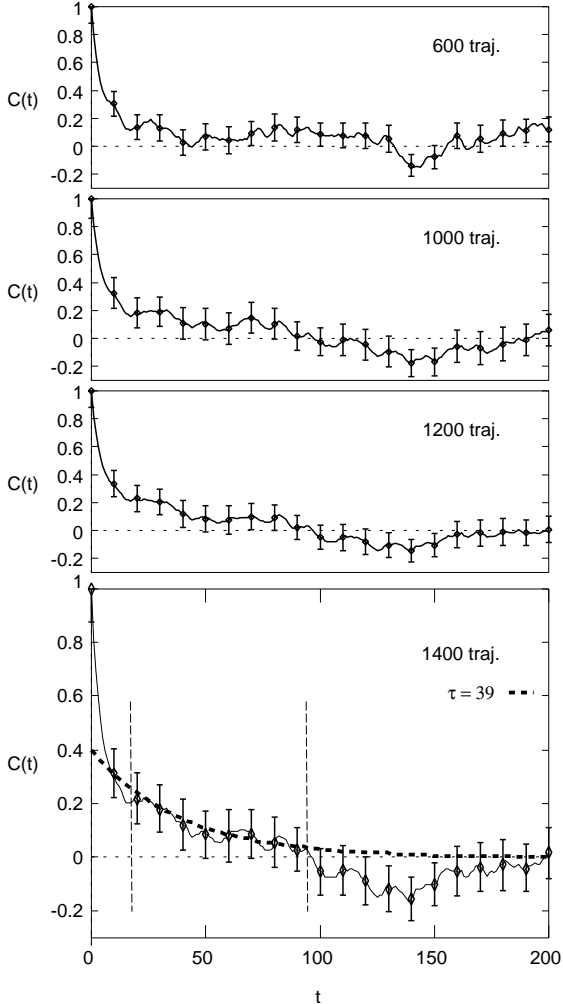}
\caption[]{
  For $\ks=0.157$, measurements on samples of increasing size are
  shown for the average plaquette.}
\label{FINITE} 
\end{figure}

We illustrate the numerical impact of finite sampling on the estimate
of autocorrelation in \fig{FINITE}.  One observes the long range
autocorrelation modes to emerge out of the noise level as the sample
is enlarged.  There is a certain threshold where the autocorrelation
function ceases to be compatible with zero and becomes visible.

Besides the average plaquette $S_\Box$, we computed the
autocorrelation of extended smeared observables that exhibit a large
ground state overlap per construction.  This corresponds to rather
long range correlations on the lattice.  We consider the Gaussian
smeared light meson masses, $m_\pi$ and $m_\rho$, and smeared spatial
Wilson and Polyakov loops built iteratively from fuzzed links of the
form:
\begin{eqnarray}
\nonumber U^0_j(x)&=&U_j(x)\\
\nonumber U_j^{n+1}(x) &=& U_j^n(x)U_j^n(x+2^n\hat j)\\
\nonumber +\sum_{|i|\neq j; i=\pm 1,\dots,3} &&
U_i^n(x)U_j^n(x+2^n\hat i) \\
&\times& U_j^n(x+2^n(\hat i +\hat j))\nonumber\\
&\times&
U^{n+}_i(x+2^{n+1}\hat j ) .
\label{W-loop}
\end{eqnarray}
$n$ labels the smearing level. We investigate these quantities for
$n=0,1,2,3$ corresponding to $1\times 1,2\times 2,4\times 4,8\times 8$
Wilson loops, respectively.

Another `fermionic' monitoring quantity is the inverse of the average
number of iterations $N_{kry}^{-1}$ of the Krylov solver.  For the
conjugate gradient (CG) it has been demonstrated in
Ref.~\cite{BLOCKED} that this quantity is related to the square root
of the ratio of the minimal to the maximal eigenvalue of a hermitian
positive definite matrix $H$, \ie, its condition number
$K=\frac{\lambda_{max}}{\lambda_{min}}$.  This is motivated by the
following considerations: let us start from the definition of $\kc$ as
the value of the hopping parameter, $\kappa$, at which the pion mass
vanishes
\begin{equation}
m_{\pi}^2 \propto m_q = \frac{1}{2} \left( \frac{1}{\kappa}
- \frac{1}{\ksc} \right) .
\end{equation}
The convergence rate of the conjugate gradient algorithm,
given a residual norm  $r$,
can be extracted from the bound \cite{ENGELI}
\begin{equation}
r \leq 2\left( \frac{\sqrt{\lambda_{max}}-\sqrt{\lambda_{min}}}{
\sqrt{\lambda_{max}}+\sqrt{\lambda_{min}}}
\right) ^{N_{CG}}.
\end{equation}
This inequality provides an estimate to the convergence behaviour of
CG on a hermitian matrix $H$.  The relation is based on the assumption
of a uniform distribution of eigenvalues of $H$.  Generally one
expects some dependence of $r$ on the detailed distribution of
eigenvalues, however.  Close to $\kappa_c$ the minimal real eigenvalue
of $M^{\dagger}M$ is small and is approximately that of $M^2$, \ie,
\begin{equation}
r \leq 2
-4 
N_{CG}
\frac{\lambda_{min}}{\lambda_{max}}
 +O\left[ 
\left(
\frac{\lambda_{min}}{\lambda_{max}}
\right)^2
\right].
\end{equation}
For the case of a poor condition number, 
$\frac{\lambda_{min}}{\lambda_{max}} \ll 1$, we can exploit this
relation and find for $r$ fixed:
\begin{equation}
N^{-1}_{CG}=\lambda\propto \frac{\lambda_{min}}{\lambda_{max}}.
\label{MINIMAL}
\end{equation}
We know empirically that the ratios of convergence rates of BiCGstab
and CG are quite constant over a rather large range of $\kappa$.  This
suggests to utilise the convergence rate of BiCGstab as an indicator
to the Monte Carlo evolution of the smallest eigenvalue of the fermion
matrix.  Since small eigenvalues correspond to large correlation
lengths, $\lambda_{min}$ presumably projects maximally onto the
slowest relevant autocorrelation mode of the system.

\begin{figure}[htb]
\epsfxsize=.465\textwidth\epsfbox{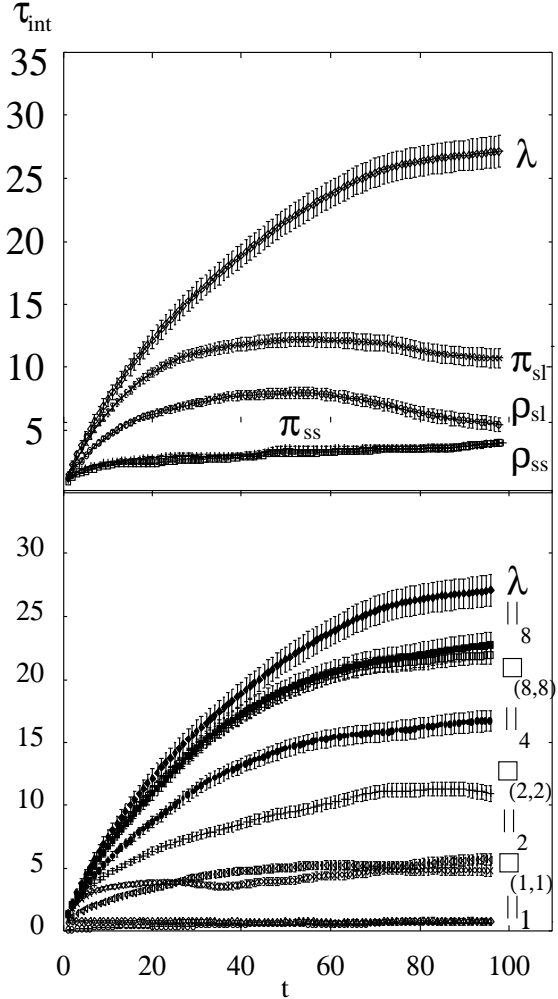}
\caption{
  For $\ks=0.157$, we demonstrate that the integrated autocorrelation
  time of $\lambda$ is an upper bound for the 
  autocorrelation times of all `fermionic' and `gluonic' observables
  measured.}
\label{HADRONGLUON} 
\end{figure}

In \fig{HADRONGLUON} the integrated autocorrelation times of sl and
ss-smeared $\pi$ and $\rho$ `masses' are shown. Here we take the
`mass' as computed from the propagator of each individual trajectory
as an estimate.  Autocorrelations of various gluonic observables
(Wilson loops and Polyakov loops) are displayed in the lower part of
the figure.  In both pictures $\lambda$, as estimated by \eq{MINIMAL},
is at the upper end for the autocorrelation times\footnote{Slower
  modes might exist for the topological charge, but seem to be
  decoupled from the observables of interest.}, together with the
$8\times 8$ Wilson loop.

The plot illustrates the quality of the signal.  It is evident that
the exponential autocorrelation times are bounded by the minimal
eigenvalue.  For all reference quantities we observed clear plateaus
in $\tau_{int}(t)$.  The Wilson and Polyakov loops provide evidence
that geometrically extended quantities indeed suffer more from
autocorrelation effects.

\begin{figure*}[p]
\epsfxsize=\textwidth\epsfbox{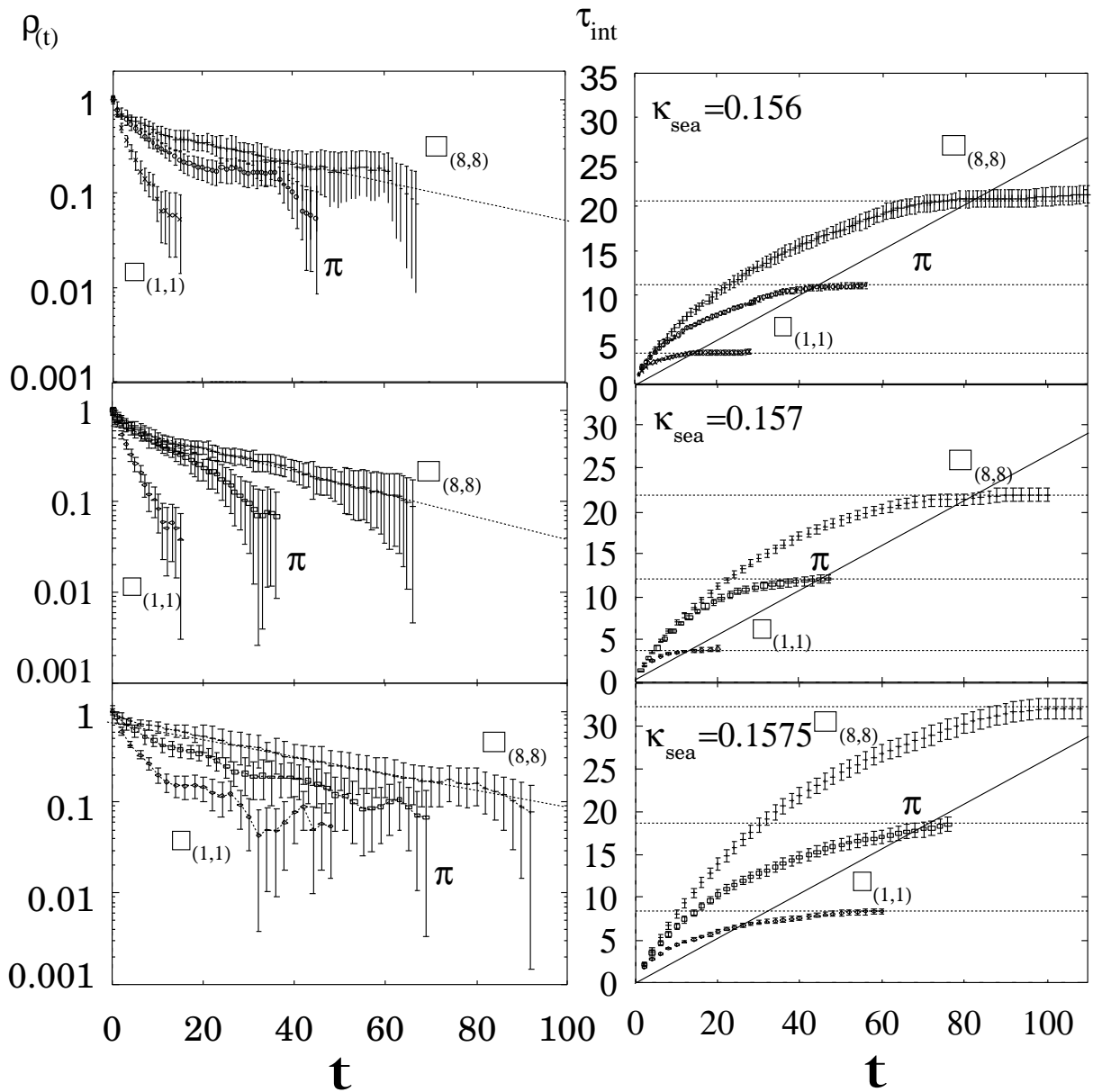}
\caption{  
  Autocorrelation functions and integrated autocorrelation times for the
  quantities: $S_\Box, \pi, \Box_{8,8}$ for all three sea quark masses. The
  2\% curve is plotted for orientation. The horizontal lines indicate the
  values of exponential and integrated autocorrelation
  times.\label{AUTODEMO}}
\end{figure*}

So far, we have compared various observables at one value of $\ks$,
$\ks=0.157$.  In the following \fig{AUTODEMO}, the sea-quark
dependency of the exponential autocorrelation function is illustrated
for three different observables: the plaquette action, the sl-smeared
pion mass and the spatial Wilson loop, $\Box_{(8,8)}$.

The smeared quantities exhibit an early asymptotic behaviour while the
un-smeared average plaquette appears to couple to many excited modes.  The
{\it integrated} autocorrelation times, $\tau_{int}(t)$, of the pion masses
and the spatial  Wilson loops as a function of the cut-off $t$ each
reach an asymptotic plateau for $t_{int}<{1/4}t$ within the estimated
errors.  Assuming a single exponential to dominate $\rho^A(t)$, the
systematic error is $\approx 2\%$ and thus smaller than the statistical
uncertainty.

\begin{figure}[htb]
{\epsfxsize=.465\textwidth\epsfbox{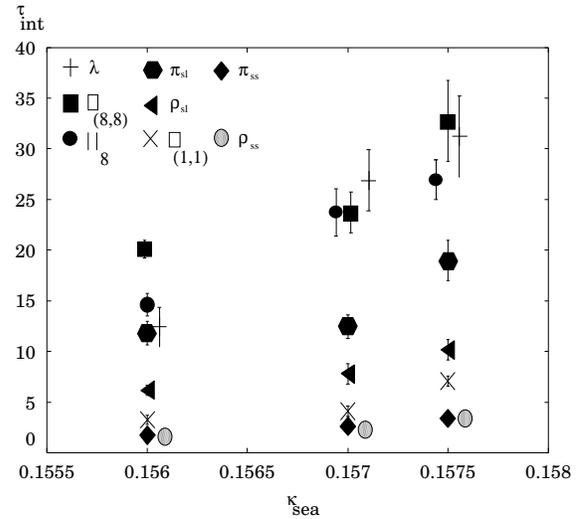}}
\caption{
  The sea quark-dependency of the estimated integrated autocorrelation
  times.
\label{AUTOALL}}
\end{figure}

A compilation of all integrated autocorrelation times can be found in 
\fig{AUTOALL}

Originally we believed that the restriction of the pseudo-fermionic
d.o.f.\ to the even sub-space \cite{ROSSI}, as offered by the o/e
preconditioning scheme for $M$, would affect the dynamics of HMC only
marginally, such that we have chosen this option.  In the LL-SSOR
scheme, however, the pseudo-fermion field has to live on the entire
lattice.

\begin{figure}[htb]
\epsfxsize=.465\textwidth\epsfbox{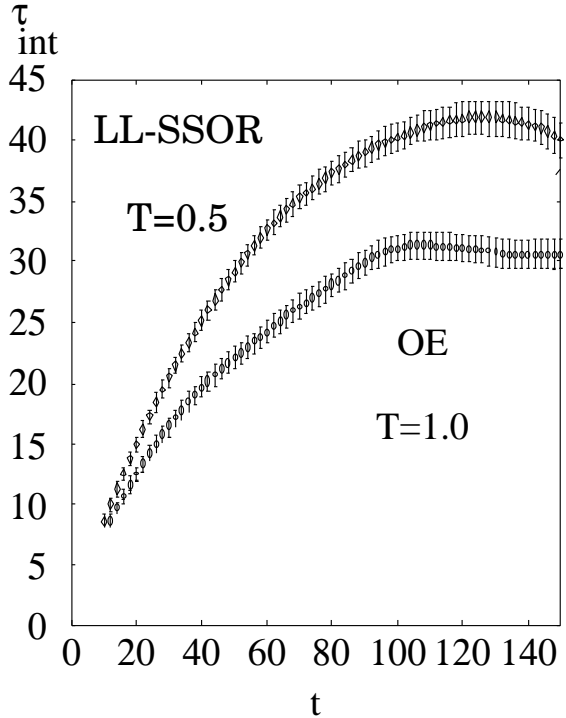}
\caption{The integrated autocorrelation time of the
  averaged number of BiCGstab iterations for $\ks=0.1575$.
\label{AUTOITER} }
\end{figure}

It is now very interesting to compare autocorrelation on the full and
the o/e reduced system, see \fig{AUTOITER}: we show the integrated
autocorrelation time of $\lambda$ for $\kappa_{sea}=0.1575$.  On the
fully occupied system, we have employed the SSOR algorithm and have
generated trajectories of half length, $T=0.5$. One would have
anticipated the autocorrelation time to increase by a factor of 2 in
trajectories. We found, however, a factor of $\sqrt{2}$ only compared
to the o/e reduced system's value.  Thus, doubling the number of
pseudo-fermionic degrees of freedom appears to affect autocorrelation
to the extent that higher stochasticity decreases the autocorrelation.

The lessons to be learnt here are: the question of optimal trajectory
length of the HMC has to be addressed anew and thinning out fermions
turns out to be counter-productive.

For the algorithmic part of this talk we conclude that the
autocorrelation times come out much smaller than anticipated
previously.  The exponential autocorrelation times all are well below
the value of $50$ trajectories.  With the knowledge of $\tau_{int}$,
we can assess the computational cost to generate {\em one}
independent full QCD vacuum state.  Taking the sustained performances
reached in the o/e and LL-SSOR preconditioned HMC (67\% and 38\%,
respectively) we estimate these numbers in Gflops-hours units, see
\tab{GFLOPS}. More detailed results will be given in \cite{SESAMAUTO}.

\begin{table*}[t]
\setlength{\tabcolsep}{.5pc}
\caption{CPU cost for the
generation of full QCD vacuum states from our Quadrics implementation.
  \label{GFLOPS}}
\begin{tabular*}{\textwidth}{@{}|c@{\extracolsep{\fill}}|c|c|c|c|c|}
\hline
\multicolumn{6}{|c|}{{\sc Lattice} $16^3 32 $ at $ \beta=5.6 $} \\ \hline
$\kappa_{sea} $&$a^{-1}_\rho$[GeV] & $L_S$[fm]&$m_\pi$ &${m_{\pi} / m_{\rho}} $-ratio & {\sc
{\sc QH2[h]} $\to $ GFlops$\times$h}  \\ \hline\hline 
0.156 (OE)    &2.19(8) & 1.44(5) &0.4482(40)& 0.8388(41) & $\approx ~ 5  \to ~\approx $ 43    \\ \hline 
0.157 (OE)    &2.25(6) & 1.39(4) &0.3412(33)& 0.7552(69)  & $\approx  22 \to ~\approx $ 189   \\ 
0.157 (SSOR)  &"   & " &"     & "          & $\approx   17 \to ~\approx $ 82    \\ \hline 
0.1575 (OE)   &2.38(7) & 1.32(4) &0.2763(29)& 0.688(12)  & $\approx  45 \to ~\approx $ 387   \\
0.1575 (SSOR) &"   & " &"     & "          & $\approx  35 \to ~\approx $ 170   \\
 \hline
\end{tabular*}
\end{table*}

\section{FLAVOUR SINGLET OPERATORS}

The computation of nucleonic matrix elements of flavour singlet
operators involves notorious noise problems when dealing with the
contributions of disconnected quark diagrams (DQD).  This presents a
headache in particular, when analysing full QCD vacuum configurations
for sea quark effects, given the limitations in the statistical
sampling.

The common technique to handle DQD is the stochastic estimator method.
\sesam\ has paid particular attention to devise methods for signal
improvements, starting out from the work of the Tsukuba and Kentucky
groups \cite{tsukuba,kentucky} in the quenched
situation \cite{disconnected}.

The most simple case of DQD to consider is the $\pi$-N $\sigma$-term,
$\sigma_N$, which amounts to determine the correlator between the
nucleon propagator, $P$, and disconnected closed quark loops,
$Tr(M^{-1})$.  According to the Kentucky technique the latter are
estimated by inverting the Dirac operator on a stochastic source, with
$Z_2$ noisy entries in all space-time and spin-colour components of
the source vector.

\begin{figure}[tb]
\epsfxsize=.465\textwidth\epsfbox[80 200 550 630]{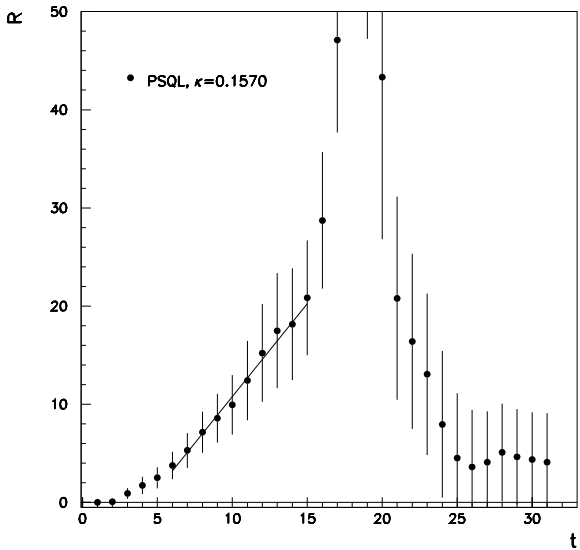}
 \caption{ \label{PLATEAU} New plateau sampling technique for the
   computation of disconnected contributions to the $\pi$-$N$ $\sigma$
   term in full QCD.}
\end{figure}

\begin{figure}[tb]
\epsfxsize=.465\textwidth\epsfbox[80 200 550 630]{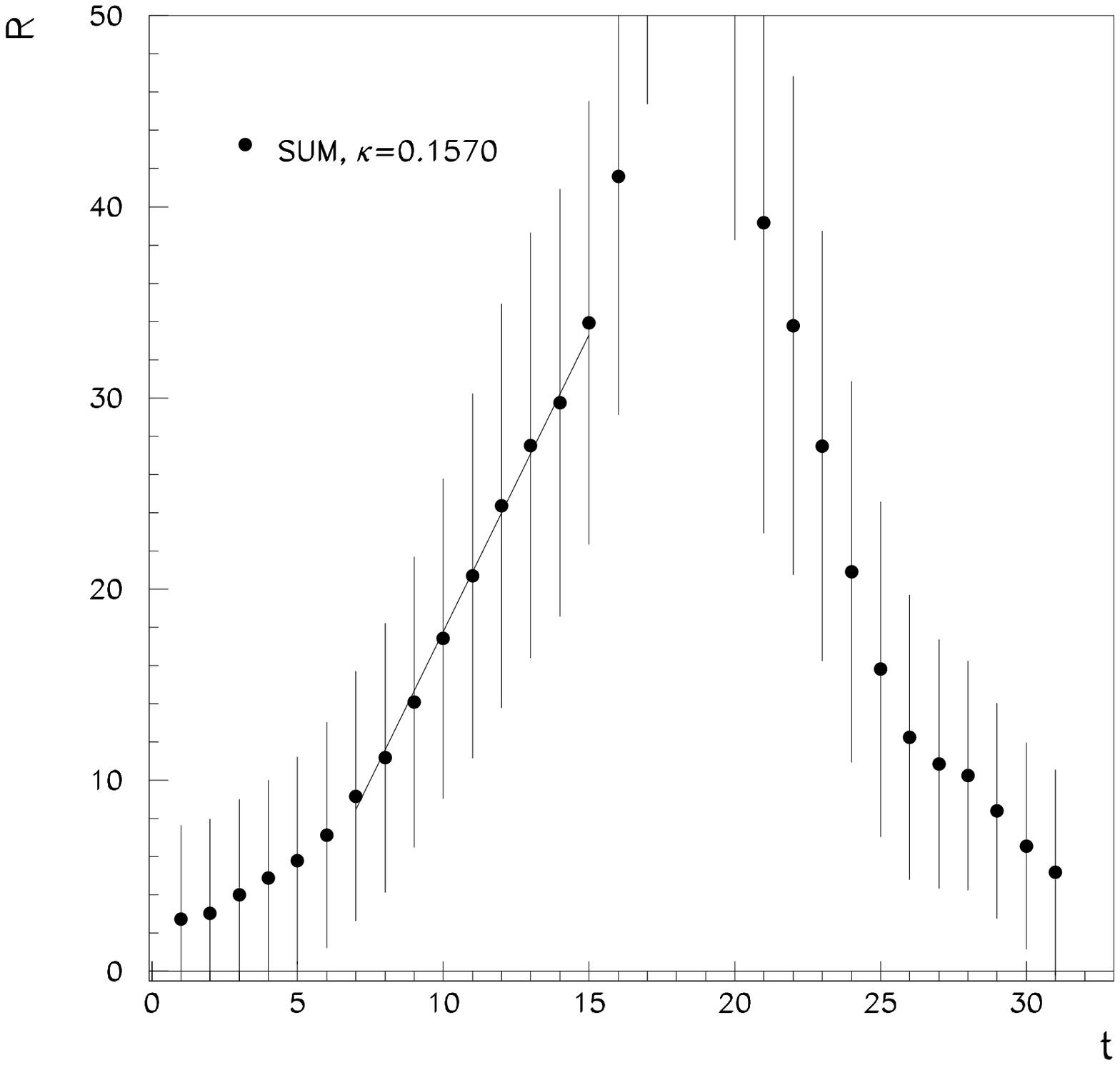}
 \caption{Same as \fig{PLATEAU}. The source is spread over the
entire lattice. \label{PLATIMPROVE}}
\end{figure}

The quantity to measure is given by the asymptotic slope in time $t$, in
the expression for the correlator
\begin{eqnarray}
\lefteqn{R(t)^{disc}}\nonumber\\
 &=& \frac{ \langle P(0 \rightarrow t) Tr(M^{-1}) \rangle}
             {\langle P(0 \rightarrow t)\rangle } - \langle Tr(M^{-1})
             \rangle \nonumber\\
&\stackrel{t \; large}{\longrightarrow}&
             \mbox{const} + t\,\langle P|\bar{q}q|P\rangle^{latt}_{disc},
\end{eqnarray}
which is prone to noise problems from the very subtraction on the
entire space-time volume. It is obvious that loops located far away
from $P(0\rightarrow t)$ in time will add in particular to this noise
level. But the highly separated terms (in time) are expected to be
decorrelated from the nucleon propagator! The obvious measure to
counteract the loss of signal is to confine the noisy source into the
time interval of $0<\tau<t$. This new {\em plateau sampling} technique
leads to a substantial improvement of the signal to $\sigma_N$, as can
be seen from \fig{PLATEAU}.
The variance turns out to be reduced by almost a factor 2, see
\fig{PLATIMPROVE}.

This result is very encouraging in view of other, more complicated
matrix elements, such as the flavour singlet axial vector current
which is of tantamount importance in proton spin studies. In this case
a weak signal has been observed on a large quenched sample by the
Tsukuba group some time ago \cite{tsukuba}.  The QCDSF collaboration,
however, could not see a signal above the noise level, within their
high statistics study of the proton structure functions \cite{qcdsf}.
It is a challenge to look for progress in the signal preparation.

\section{LIGHT HADRON SPECTRUM\label{SPECTRUM}}

With three \sesam\ samples at $\beta=5.6$ (and one \tkl\ sample at
$\ks=0.1575$) we can perform the chiral extrapolation in $\ks$ for hadronic
observables.  We try to estimate the influence of dynamical fermions by
comparison with quenched QCD at the equivalent scale.  At each sea-quark
sample zero-momentum two-point functions for mesons and baryons ({\em cf.}
\tab{OPERATORS}) are computed for various valence hopping parameters $\kv$.
\begin{table}[htb]
\setlength{\tabcolsep}{.9pc}
\caption{Hadron operators.\label{OPERATORS}}
\begin{tabular}{|l|c|}
\hline
meson        &   $\chi(x)$ \\ 
\hline\hline
pseudoscalar &   $\chi_{PS}(x) = \bar{q}'(x) \gamma^5 q(x)$\\
vector       &   $\chi^{\mu}_{V}(x) = \bar{q}'(x) \gamma^{\mu} q(x)$\\
scalar       &   $\chi_{Sc}(x) = \bar{q}'(x) q(x)$\\
axial-vector  &   $\chi_{Ax}(x) = \bar{q}'(x) \gamma_5 \gamma^{\mu} q(x)$\\
\hline \hline
baryon       &   $\chi(x)$ \\ \hline
nucleon      & $\chi_{N}(x) = \epsilon _{abc} ( q_a C \gamma_5 q_b) q_c$\\
$\Delta$     & $\chi^{\mu}_{\Delta}(x) = \epsilon _{abc} ( q_a  
C \gamma^{\mu} q_b) q_c$\\ \hline
\end{tabular}
\end{table}
Altogether we work with fifteen mass
estimates per $\ks$, see \tab{kappas}.
\begin{table}[htb]
\setlength{\tabcolsep}{.5pc}
\caption{Valence kappa values.\label{kappas}}
\begin{tabular}{|c|c|}
\hline
$\ks$   & $ \{ \kv \}$ \\ \hline\hline
0.156 & $\{0.156, 0.157, 0.1575, 0.158, 0.1585 \}$ \\ \hline
0.157 & $\{ 0.1555, 0,156, 0,1565, 0.157, 0.1575 \}$ \\ \hline
0.1575 & $\{ 0.1555, 0,156, 0,1565, 0.157, 0.1575 \}$ \\ 
\hline
\end{tabular}
\end{table}

We want the propagators to be dominated by the lightest state at small
time separations. Therefore we apply Wuppertal smearing with smearing
parameter $\alpha = 4$ and 50 smearing iterations.  Smeared-smeared
(ss) and smeared-local (sl) correlators are used in simultaneous
mass-estimates. We fit to the data on time-slices 10 to 15 (see
Ref.~\cite{SPECTRUMPAPER} on how to determine the fit ranges and all
that).  Autocorrelation times of $\tau_{\rm int}<25$ trajectories
have suggested to calculate propagators on configurations separated by 25
HMC trajectories, the residual correlation is taken into account by
jackknife binning.

We perform single-exponential fits:
\begin{eqnarray}
\label{single_ex}
&&C(t)_{\rm{mes}}  =  A ( e^{-m t} + e^{-m (T-t)} ), \nonumber \\
&&C(t)_{\rm{bar}}  =  A e^{-m t} .
\end{eqnarray}
The effective masses are determined iteratively for the mesons, by solving
the equation
\begin{eqnarray}
&&\frac{C_{AB}(t)}{C_{AB}(t+1)} \nonumber\\
&&= \frac{e^{-m_{\rm{eff}}(t) t} + e^{-m_{\rm{eff}}(t) (T-t)}}
       {e^{-m_{\rm{eff}}(t) (t+1)} + e^{-m_{\rm{eff}}(t) (T-t-1)}},
\end{eqnarray}
and directly for  baryons,
\begin{equation}
m_{\rm{eff}}(t) = \rm{log} \frac{C_{AB}(t)}{C_{AB}(t+1)}.
\end{equation}

\subsubsection*{Light Mesons}
\begin{figure}[htb]
\epsfxsize=.465\textwidth\epsfbox{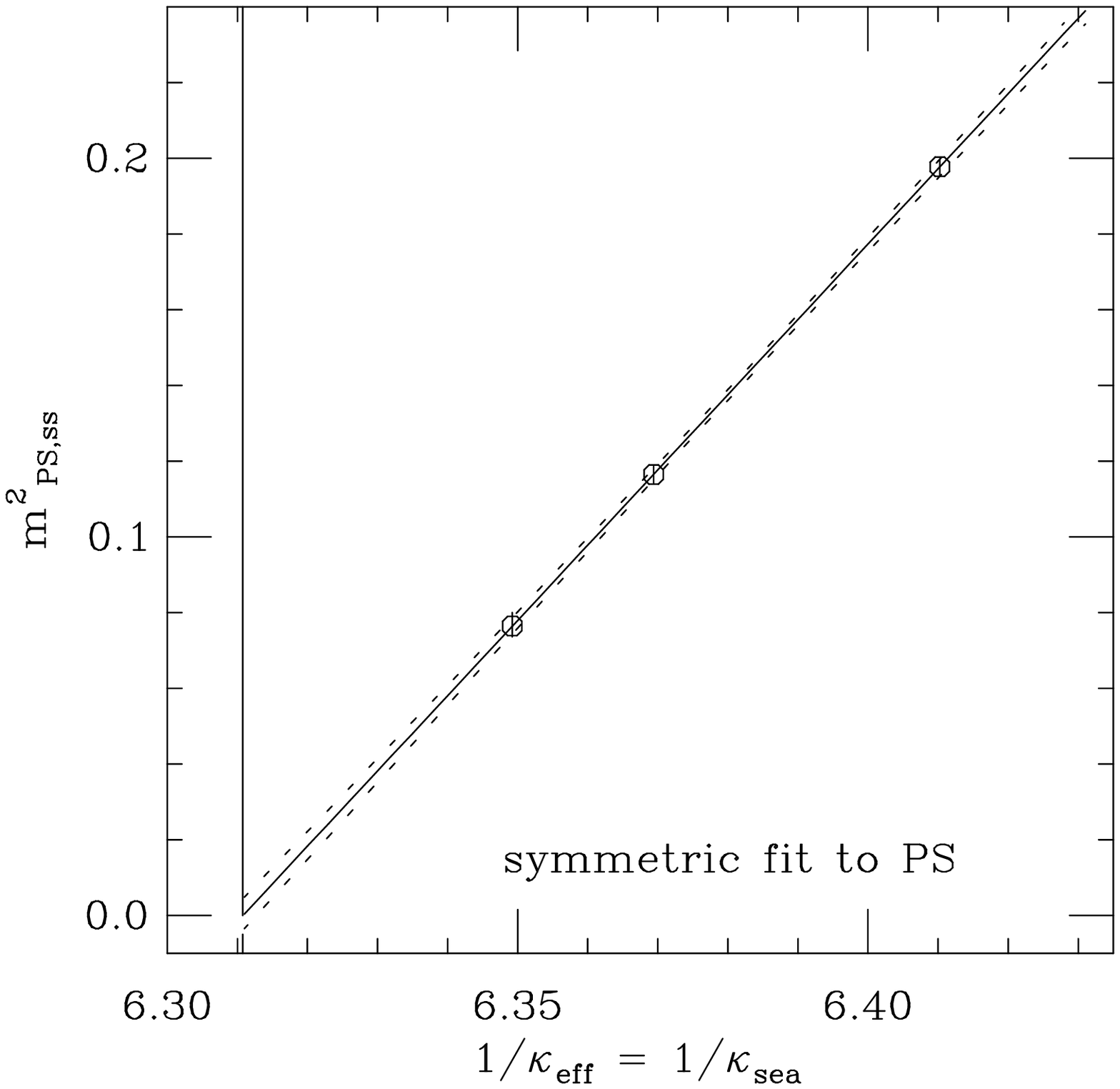}
\epsfxsize=.465\textwidth\epsfbox{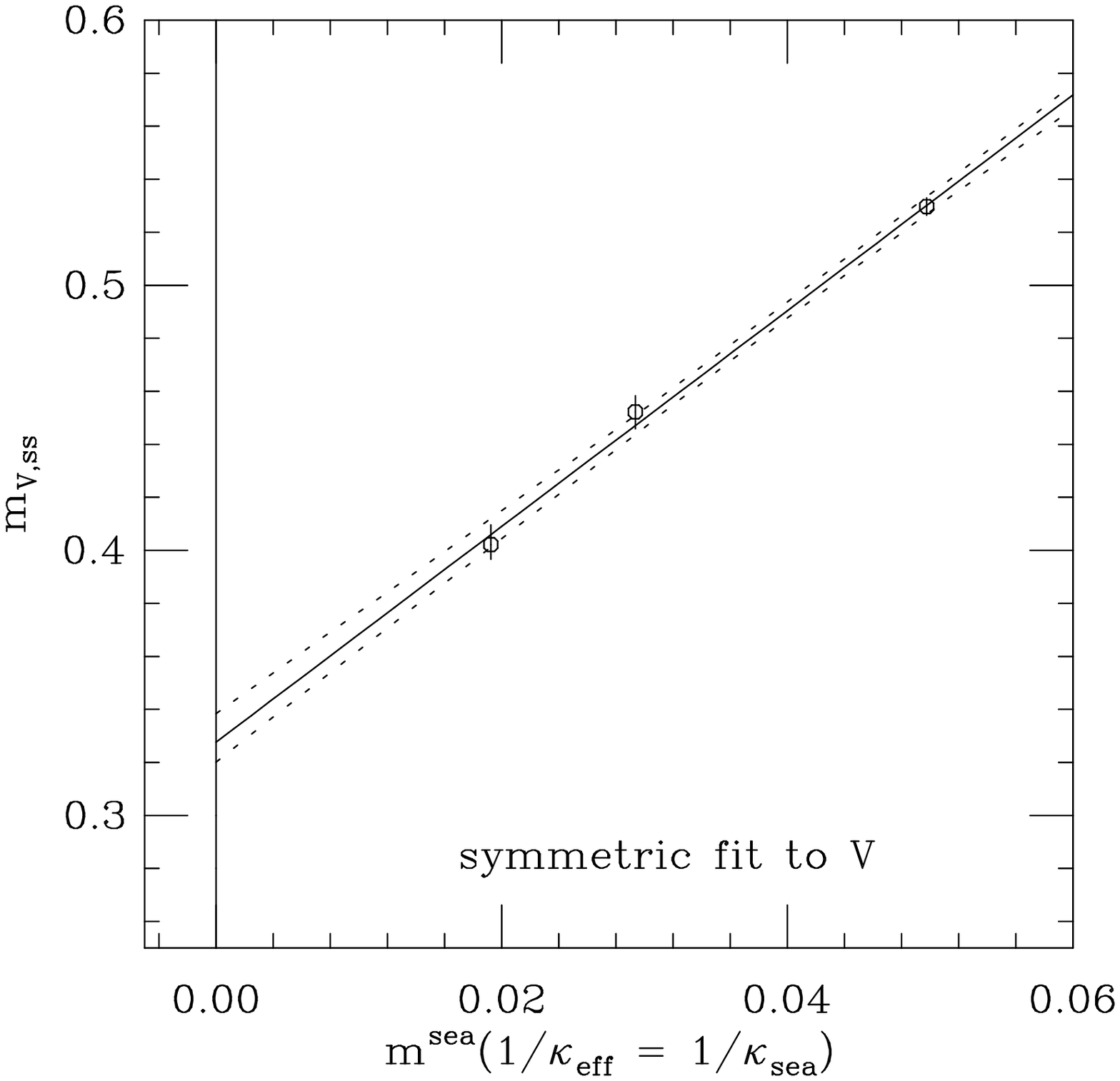}
\caption{$\mpsot$ as a function of $\oks$ and $\mvot$ as a
  function of $\mqs$ (in lattice units).
\label{MASS1}}
\end{figure}

We carry out linear extrapolaions in the lattice quark mass, using data
with $\kv=\ks$, generically called $\mot$:
\begin{eqnarray}
&&  \mpsot = c \left( \oks - \okc \right)\; , \label{pisym} \\ 
&& \mvot =  m^{\rm crit} + b \mpsot \; .\nonumber
\end{eqnarray}
These fits, called ``symmetric'', are shown in \fig{MASS1}. We
find the pseudoscalar mass to be  well matched by the linear
ansatz (with a $\chisq = 0.002$), whereas the vector masses may exhibit
some downward curvature (based on a $\chisq = 1.1$). 
The lattice spacing is obtained from $M_\rho$ as physical
input\footnote{Continuum masses go with capital ``$M$'', whereas
  lattice masses are written as ``$m$''.}.  The results are quoted at both
the chiral limit and  the experimentally given mass ratio, defining
$\ksl$:
\begin{equation}
\mpr=0.1785.\label{MPR}
\end{equation}

The lattice spacings from the $\rho$ become
\begin{eqnarray}
&&a^{-1}_{\rho} = 2.35(6)\;{\sf GeV} \;\;\;\;{\rm at }\;\; \ksc  , \nonumber\\
&&a^{-1}_{\rho} = 2.33(6)\;{\sf GeV} \;\;\;\;{\rm at }\;\; \ksl,
\end{eqnarray}
with
\begin{equation}
\ksc = 0.15846(5),\;\;\;\;  \ksl = 0.15841(5) \label{res1} .
\end{equation}

\subsubsection*{Strange Mesons}

For want of 3 dynamical quarks, we have to find a way to compute
strange mesons in a sea of two light mesons. This requires valence
quarks different from sea quarks.  We define an effective $\kappa$
through $ {\frac{1}{\kappa^{\rm eff}_{\rm v}}} = \frac{1}{2} \left(
  \frac{1}{\kappa^1_{\rm v}} + \frac{1}{\kappa^2_{\rm v}} \right)$,
where $\kappa^1_{\rm v}$ and $\kappa^2_{\rm v}$ refer to valence
quarks in a meson.  

In Ref.~\cite{SESAMLIGHT}, we have introduced a notation to specify these
valence quarks with masses different from sea quark masses.  $\mot$ refers
to both $\mqv$'s equal $\mqs$ of the sample, $\moth$ means only one $\mqv$
equal to $\mqs$, and $\mtf$ stands for neither $\mqv$ equal to $\mqs$.  We
argued that $\mpstf$ and $\mpsoth$ as linear functions of $\oks$ and $\okv$
can be parametrised by the two sets of equations:
\begin{eqnarray}
&&\left( 
\begin{array}{c} 
\mpsot  \\
\mpsoth \\
\mpstf  
\end{array}
\right) =  \nonumber\\
&&\left( 
\begin{array}{cc} 
2 c & 0  \\
c + c_{13} & c - c_{13} \\
c_{34} & 2c - c_{34} 
\end{array}
\right)   
\left( 
\begin{array}{c} 
\mqs \\
\mqv
\end{array}
\right),
\label{quarkmatrix}
\end{eqnarray}
\begin{equation}
\left( 
\begin{array}{c} 
\mvot  \\
\mvoth \\
\mvtf  
\end{array}
\right) =  m^{\rm crit} \, + \, b 
\left( 
\begin{array}{c} 
\mpsot  \\
\mpsoth \\
\mpstf  
\end{array}
\right),
\label{quarkmatrix1}
\end{equation}
with $c_{34} = 2 c_3$ and $c_{13} = 2c_3' - c$ and $ \mqv = \frac{1}{
  2}\left(\frac{1}{\kappa_{\rm v}^{\rm eff}} - \okc \right)$.  A combined
linear fit of all the pseudoscalar data with the ansatz of \eq{quarkmatrix}
leads to an acceptable $\chisq = 4.4/23$. For further details and the
vector meson fits we refer to \cite{SPECTRUMPAPER}.

We determine $\kst$ from $\moth$ by matching
\begin{equation}
\frac{m_{\rm V, {\sf sv}}(\ksl,\kst)}{
 m_{\rm V, {\sf ss}}(\ksl)} = \frac{M_{K^*}}{M_{\rho}} = 1.16 \; ,\label{MKSR}
\end{equation}
where $\kl$ is given by \eq{res1}.  Alternatively, $\kst$ is computed
from $\mtf$ matching
\begin{equation}
\frac{m_{V_{1}, {\rm vv}}(\ksl,\kst)}{ m_{V_{2}, {\rm ss}}(\ksl)} =
\frac{M_{\phi}}{M_{\rho}} = 1.326 \; .\label{MPHIR}
\end{equation}

In this manner we can calculate the masses of the $K$ ($K^*$) and the
$\phi$ composed of the appropriate light and strange quarks in a sea of
light quarks.

\subsubsection*{Baryon Masses}
The remaining independent quantities to determine after fixing $\kc$
via pseudoscalar meson mass and $a$ via $\rho$ mass are the masses of
nucleon and $\Delta$. We perform the chiral extrapolation by fitting
to a linear function, the systematic error is taken from quadratic 3
parameter fits.

\begin{figure}[htb]
\epsfxsize=.465\textwidth\epsfbox{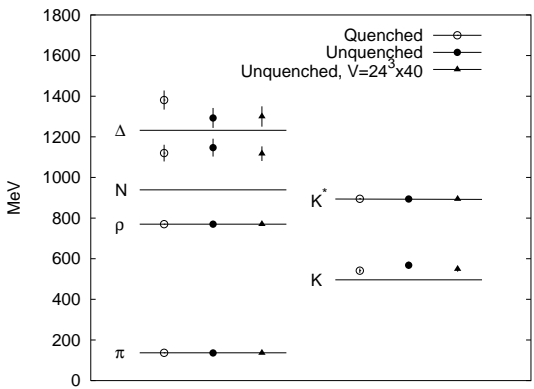}
\caption{The light mass spectrum in the `simultaneous' chiral limit.
\label{BARYON}}
\end{figure}

In \fig{BARYON} we show an overview of light and strange hadron masses
comparing our three simulations (\sesam, \tkl, and quenched)\footnote{
  The \tkl\ data are preliminary ($\kappa=0.1575$ only).}.  $\pi$ and
$\rho$ are used to fix scales and chiral limit and therefore must
coincide with experimental values. For the $\Delta$, sea quark effects
seem to be visible, while for the nucleon there is no observable
difference to quenched simulations.  It remains to be seen how these
findings are affected including the $\kappa=0.158$ sample from \tkl.
As expected finite volume effects (comparing extrapolations on \sesam\ 
data with \tkl\ data for $\kappa=0.1575$) are largest for nucleon and
$\Delta$ (around 5 to 8 \% ).  The conversion to physical units
slightly softens this effect since the $\rho$ is some 2 to 3 \%
lighter on the $24^3$ lattices.
From the statistical accuracy achieved a sensible continuum
extrapolation appears to be in reach of Tera-computer performance!

\section{LIGHT QUARK MASSES}

Light quark masses--albeit being fundamental parameters of the
Standard Model--are not directly accessible to experiment.

In principle, lattice methods allow to compute their absolute values
from the QCD Lagrangian, using the values of hadron masses as physical
input \cite{APELU,mackenzie,gupta}. Wilson fermions appear appropriate
here as--unlike staggered fermions--they discretise full QCD with
correct flavour attributes.

In a recent analysis of world data, unquenching (from $N_f =0$ to $N_f
= 2$) seems to lower the values of light and strange quark masses by
about 50 \% \cite{gupta}.  To clarify these indications we have
investigated, within the \sesam\ project, the mass renormalization of
the light and strange quarks under the influence of two dynamical
flavors \cite{SESAMLIGHT}.

In order to extract the masses of the light and the strange quark from
our meson data the lattice results have been extrapolated to the
experimental mass ratios \eq{MPR}, \eq{MPHIR}, and \eq{MKSR}.
At each sea-quark sample we evaluate meson masses
with strange valence quarks and perform an extrapolation of these
masses to the physical sea of light quarks. 

Under the condition that we have to treat the $u$-$d$ isospin doublet
as degenerate we can extract the masses of the degenerate light quarks
in a sea of degenerate light quarks as well as the mass of the strange
quark.

We take the  values from our light spectrum computations 
and  convert to physical units in the
$\overline{MS}$-scheme according to:
\begin{equation}
m_{\overline{MS}}(\mu) = Z_M(\mu a) \mqs a^{-1}_{\rho} \; ,
\end{equation}
$Z_M(\mu a)$ is computed in boosted 1-loop perturbation theory
\cite{APELU,Lepage}. Finally we run the values to 2 ${\sf GeV}$,
see \tab{LIGHTMASSES}.

\begin{table*}[tb]
\setlength{\tabcolsep}{1.5pc}
\caption{$\kappa$-values and corresponding light quark masses. The
  scale is taken from the $\rho$: $a_{\rho}^{-1} 2.33(6)$ GeV at $\ksl$.}
\label{LIGHTMASSES}
\begin{tabular*}{\textwidth}{@{}|l@{\extracolsep{\fill}}|l|r|}
\hline
        & $\kappa$     & $m_{\overline{MS}}(2\, {\sf GeV})$ \\
\hline\hline
\multicolumn{3}{|c|}{$N_f=2$}\\
\hline\hline
light   & 0.15841(5)   & 2.7(2) MeV                         \\
strange & 0.15615(20)$^{\mbox{stat}}$(20)$^{\mbox{syst}}$ & 140(20) MeV \\
\hline
strange sea$\,\,\,\,\,\,\,\,$ & 0.15709(12) & 80(8) MeV \\
\hline\hline
\multicolumn{3}{|c|}{$N_f=0$}\\
\hline\hline
light   & & 5.5(5) MeV \\
strange & & 166(15) MeV \\
\hline
\end{tabular*}
\end{table*}

We remark that one could have determined $\kst$ by matching the ratio
$\frac{M_{\phi}}{M_{\rho}}$ using the symmetric fit only \cite{gupta},
see \tab{LIGHTMASSES}.  This result, $m^{\rm
  strange}_{\overline{MS}}(2\,{\sf GeV}) = 80(8)\, {\sf MeV}$ turns
out to be much smaller than the value found before, $m^{\rm
  strange}_{\overline{MS}}(2\,{\sf GeV}) = 140(20)\, {\sf MeV}$.
However, the $\phi$ would consist of strange valence quarks under the
influence of a sea of strange quarks, which is a poor description of
the physically correct situation of two light dynamical quarks.

Comparing our results to the quenched values at corresponding
$\beta_{\rm quenched} = 6.0$, we observe a much smaller dynamical
light quark mass \cite{gupta}.  $m^{\rm strange}/m^{\rm light} \approx
52$, whereas the strange mass is compatible to the quenched value
within errors.

Let us comment on the dramatic change in the light quark mass due to
unquenching. We can define a quark mass at {\em fixed} sea-quark:
\begin{equation} 
\mqv = \frac{1}{
  2} (\frac{1}{\kappa_{\rm v}^{\rm eff}} - \okvc) \; .
\label{qmass1}
\end{equation}
Setting $\mpstf (\kvc) = 0$ at $\kvc \neq \ksc$ forces the `pion' to become
massless at $\kvc$.

\begin{figure}[htb]
\epsfxsize=.465\textwidth\epsfbox{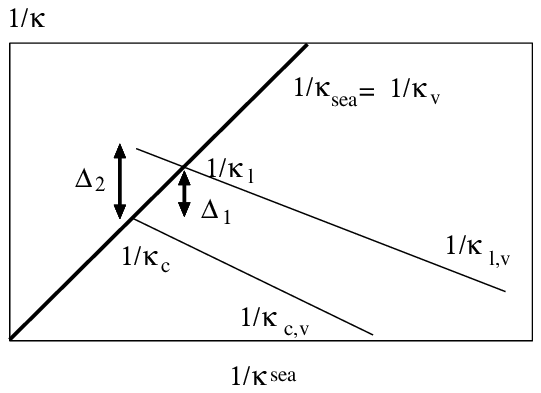}
\caption{Schematical
  plot of $\okvc$ and $\okvl$ vs.\ $\oks$.
\label{geom}}
\end{figure}

In the manner of quenched computations we measure a `bare light quark
mass' at each of the three sea-quark values.  A chiral extrapolation
of these quark masses in the sea-quark to the chiral point would yield
the value $\Delta_2$ in \fig{geom}, while the true value (from a pion
at a physical sea-quark) is given by $\Delta_1 < \Delta_{2}$.  We
might try to solve this $\Delta_{1}$-$\Delta_2$-discrepancy and
extrapolate $\Delta_2$ to the location of the light sea-quarks.
However, either we give up working at the physical pion mass---as the
critical kappa $\okvc$ would become too low otherwise---or again the
`quark mass' given in this way is too large by about a factor 2.
As these `light quark masses' are very similar to that of the quenched
simulation ($5.7(4)$, $5.6(3)$, $5.4(3)$ {\sf MeV}), the conclusion is
that it is not possible to estimate the light quark mass from quenched
computations since the light quark mass known by symmetric
extrapolation cannot be recovered computing in valence quark style at
fixed sea quark mass, with subsequent extrapolation in $\mqs$.

\section{HEAVY QUARKONIA}

As has been reported by C. Davies in her contribution to this
workshop, nonrelativistic QCD (NRQCD), an effective theory for the low
energy regime of heavy quarkonia, has proven to be an efficient tool
to directly calculate bottomonium on the lattice \cite{1}.  Full QCD
simulations using dynamical staggered quarks have shown the
sensitivity of fine and hyperfine splittings to vacuum polarisation.
Lattice observables extrapolated to $N_f=3$ have turned out to be in
remarkable agreement with experimentally known quantities, thus
unknown quantities can be predicted with some confidence.

In this section, we present preliminary results of our systematic
study of lattice NRQCD for $b\bar b$ systems with dynamical Wilson
quarks at three different masses, an NRQCD action that includes
relativistic corrections of order ${\cal O}(M_bv^6)$, mean field
improvement with $u_0$ computed from the mean link in Landau gauge and
efficient wave function smearing.

\subsection{Technique of NRQCD}

The nonrelativistic Lagrangian is decomposed into the kinetic energy operator,
\begin{equation}
  H_0 = -\frac{\Delta^{(2)}}{2M_b} \; ,
\end{equation}
which is of order $M_bv^2$, and relativistic corrections the
importance of which is estimated via power counting. We include
operators of order $M_bv^4$ and $M_bv^6$ \cite{2},
\begin{equation}
  \delta H = \delta H^{(4)} + \delta H^{(6)} \; ,
\end{equation}
with
\begin{eqnarray}
 \delta H^{(4)} =\nonumber\\
  &-& c_1\frac{\left(\Delta^{(2)}\right)^2}{8M_b^3} \nonumber\\
  &+& c_2\frac{ig}{8M_b^2}\left(\boldmath\Delta\cdot E -
  E\cdot\boldmath\Delta\right) \nonumber \\
  &-& c_3 \frac{g}{8M_b^2}\sigma\cdot\left( \tilde\Delta\times\tilde {\bf E} -
  \tilde{\bf E}\times\tilde\Delta\right)\nonumber\\
  &-& c_4\frac{g}{2M_b}\sigma\cdot
  \tilde{\bf B} \nonumber \\
  &-& c_5 \frac{a^2\Delta^{(4)}}{24M_b} - c_6
  \frac{a\left(\Delta^{(2)}\right)^2}{16nM_b^2}, 
\end{eqnarray}
and
\begin{eqnarray}
\lefteqn{  \delta H^{(6)} =
  - c_7 \frac{g}{8M_b^3}\lbrace
  \Delta^{(2)},\sigma\cdot {\bf B}\rbrace} \nonumber \\
  && - c_8 \frac{3g}{64M_b^4} \lbrace
  \Delta^{(2)},\sigma\cdot\left(\Delta\times{\bf E} - {\bf
  E}\times\Delta\right)\rbrace \nonumber \\
  && - c_9 \frac{ig^2}{8M_b^3}\sigma\cdot{\bf E}\times {\bf E} \; .
\end{eqnarray}

Derivatives and fields with tilde have their leading order
discretisation errors removed in order to correct for ${\cal O}\left(
  a^2M_bv^4\right)$ errors.  Following Ref.~\cite{1} the quark Green's
function is calculated from the evolution equation
\begin{eqnarray}
  G(t+1) & = & \left( 1 - \frac{aH_0}{2n}^n\right)U_4^{\dag}\left( 1 -
  \frac{aH_0}{2n}\right)^n\nonumber\\
         &\times& \left( 1 - a\delta H \right) G(t), \\
  G(0) & = & \delta_{{\bf x},0} \; .
\end{eqnarray}
The parameter $n$ allows to stabilise the evolution in case of small
bare quark masses. For the $\Upsilon$ system $n=2$ is appropriate. The
Lagrangian is tadpole improved a fact that may justify a tree level
matching to QCD, i.e. all the coefficients $c_i$ are set to one. Note,
however, that first order perturbative corrections to some
interactions may well be of the same sizes as relativistic ${\cal
  O}(M_bv^6)$ corrections. We choose $u_0$ to be the mean link in
Landau gauge as recently there have been hints for better scaling
properties associated with this choice compared to the average
plaquette prescription \cite{3}. We find a two percent difference in
$u_0$ between both choices for all $\kappa$ values. The tadpole
improved chromo-fields then differ by about eight percent. Through the
whole simulation we fix the heavy quark mass to a value $aM_b = 1.7$.
Taking advantage of the smallness of the bottomonium system, we
exploit configurations more than once by starting the propagator
evolution both at different spatial source points located on one and
the same timeslice and on different timeslices.

Meson correlation functions are built from quark propagators combined
with suitable interpolating operators:
\begin{eqnarray}
\lefteqn{  G^{\rm meson}_{\rm sc,sk}\left( t\right) = }\nonumber\\
&&\sum_{\bf x,y} {\rm Tr} \left[
  G^{\dag}\left({\bf x},t\right)\Gamma_{(sk)}^{\dag}\left({\bf y -
  x}\right)\tilde G\left({\bf y},t \right)\right] \; , 
\end{eqnarray}
where the source smeared propagator $\tilde G$ is obtained by evolving
the extended source:
\begin{equation}
  \tilde G\left({\bf y}\right) = \sum_{\bf x} G\left({\bf y -
  x},t\right) \Gamma_{(sc)}\left({\bf x}\right) \; .
\end{equation}
We adopt spectroscopic notation and denote (radially excited)
spin-parity eigenstates by $n^{2S+1}L_J$. The interpolating operator
$\Gamma^{(sc/sk)}({\bf x}) \equiv \Omega \Phi^{(sc/sk)}\left({\bf
    x}\right)$ contains a spin matrix and a spatial smearing function.
The latter is calculated as the solution of the Schr\"odinger Equation
with the Cornell potential for definite radial quantum number and
angular momentum. Note that 'local' P-states are realized through
derivatives acting on the delta function. For the $\Upsilon$ and
$\eta_b$ we calculate a $4\times 4$ matrix of correlations with $sc/sk
= l,1,2,3 $ corresponding to a point source, the ground state, the
first and second excited states respectively. For the $L=1$ states we
restrict ourselves to the ground state and the first excitation as
signals are worse. Gauge configurations are fixed to Coulomb gauge.

\subsection{Data Analysis}

\begin{figure}[p]
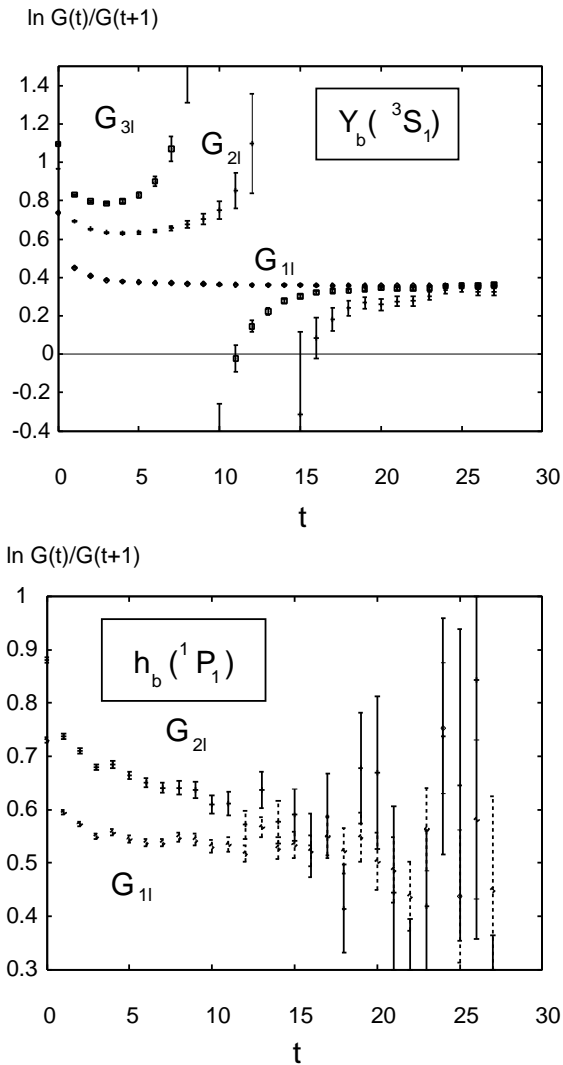

\epsfxsize=.465\textwidth\epsfbox{effmass3.eps}
\epsfxsize=.465\textwidth\epsfbox{effmass1.eps}
 \caption{Effective masses of smeared-local correlators for the
 \label{LM}
 $\Upsilon$ and $h_b$, $\kappa_{sea} = 0.1575$, $n_{config} = 700$.
 Propagator indices label the radial quantum number of the smearing
 function. Note that $G_{2l}$ and $G_{3l}$ rise sharply indicating the
 sudden decay of the dominating excited state to the ground state.}
\end{figure}

Figure \ref{LM} gives an impression of the signals' quality.
Concerning the $L=0$ states we are able to force correlators to stay
in the first excitation for about ten time-steps. Local masses for P
states are much noisier and drop to the ground state without dwelling
in an 'excited plateau' first. To extract energies we fit several
correlators simultaneously to a multi-exponential ansatz. We find that
vector fits to smeared-local propagators,
\begin{equation}
  G_{\rm meson}^{(sc,l)} (t) = \sum_{k=1}^{n_{exp}} b_{sc,k} {\rm
    e}^{-E_kt}\; ,
\end{equation}
are quite stable whereas matrix fits demand for higher statistics.
Tab.~\ref{tab:fits} gives a representative sample of fits.  We
determine hyperfine splittings by single exponential fits to the ratio
of smeared-local propagators thus exploiting the strong correlation
between them. More complicated fit-functions confirm the results
obtained from the single exponential ansatz but do not behave very
stable.
\begin{table*}[t]
\setlength{\tabcolsep}{1.pc}
\caption{Vector Fits to smeared-local correlators for $\Upsilon$ and
  $h_b$ states. Errors are calculated from 200 bootstraps\label{tab:fits}.}
\begin{tabular*}{\textwidth}{@{}|c@{\extracolsep{\fill}}|c|c|c|c|c|c|c|}
\hline 
$N_{exp}$ & $t_{min}$ & $t_{max}$ & $aE_{1}$ & $aE_{2}$ & $aE_{3}$ &
$\chi^2 / dof $ & Q  \\ \hline\hline 
\multicolumn{8}{|c|}{{$\bf ^3S_1$ }}\\
\hline 
$ 2 $ & $ 4 $ & $ 30 $ & $ 0.3589(8) $ & $ 0.605(6) $ &$ - $ &$ 54.5 / 48 $ & $  0.241 $ \\ 
& $ 6 $ & $ 30 $ &$ 0.3587(8) $ &$ 0.599(7) $ & $ - $ & $ 46.9 / 44 $ & $  0.355 $ \\ 
& $ 8 $ & $ 30 $ &$ 0.3586(8) $ &$ 0.599(10)$ & $ - $ & $ 46.1 / 40 $ & $  0.235 $ \\ 
& $ 10 $ & $ 30 $ &$ 0.3583(9)$ &$ 0.585(15)$ & $ - $ & $ 44.1 / 36 $ & $  0.167 $ \\ 
\hline 
$ 3 $ & $ 4 $ & $ 25 $ & $ 0.3578(9) $ &$ 0.589(12) $ &$ 0.782(18) $ &$ 63.0 / 54 $ & $  0.188 $ \\ 
& $ 5 $ & $ 25 $ &$ 0.3581(10)$ &$ 0.592(13) $ &$ 0.762(28) $ & $ 61.2 / 51 $ & $  0.155 $ \\ 
& $ 6 $ & $ 25 $ &$ 0.3577(12)$ &$ 0.574(18)$ &$ 0.741(32) $ & $ 58.2 / 48 $ & $  0.148 $ \\ 
& $ 7 $ & $ 25 $ &$ 0.3579(11)$ &$ 0.581(20)$ &$ 0.78(5) $ & $ 56.0 / 45 $ & $  0.126 $ \\ 
\hline
\multicolumn{8}{|c|}{{$\bf ^1P_1$ }}\\
\hline
$ 2 $ & $ 4 $ & $ 30 $ & $ 0.533(5)$ &$ 0.728(20) $ &$ - $ &$ 48.4 / 48 $ & $  0.457 $ \\ 
& $ 5 $ & $ 30 $ & $ 0.535(6)$ & $ 0.724(15) $ & $ - $ & $ 41.2 / 46 $ & $  0.672 $ \\ 
& $ 6 $ & $ 30 $ &$ 0.536(8)$ & $ 0.719(21)$ & $ - $ & $ 41.1 / 44 $ & $  0.597 $ \\ 
& $ 7 $ & $ 30 $ &$ 0.533(10)$ &$ 0.713(28)$ & $ - $ & $ 40.6 / 42 $ & $  0.534 $ \\ 
\hline
\end{tabular*} 
\end{table*}

\subsection{Results}

\begin{figure*}[htb]
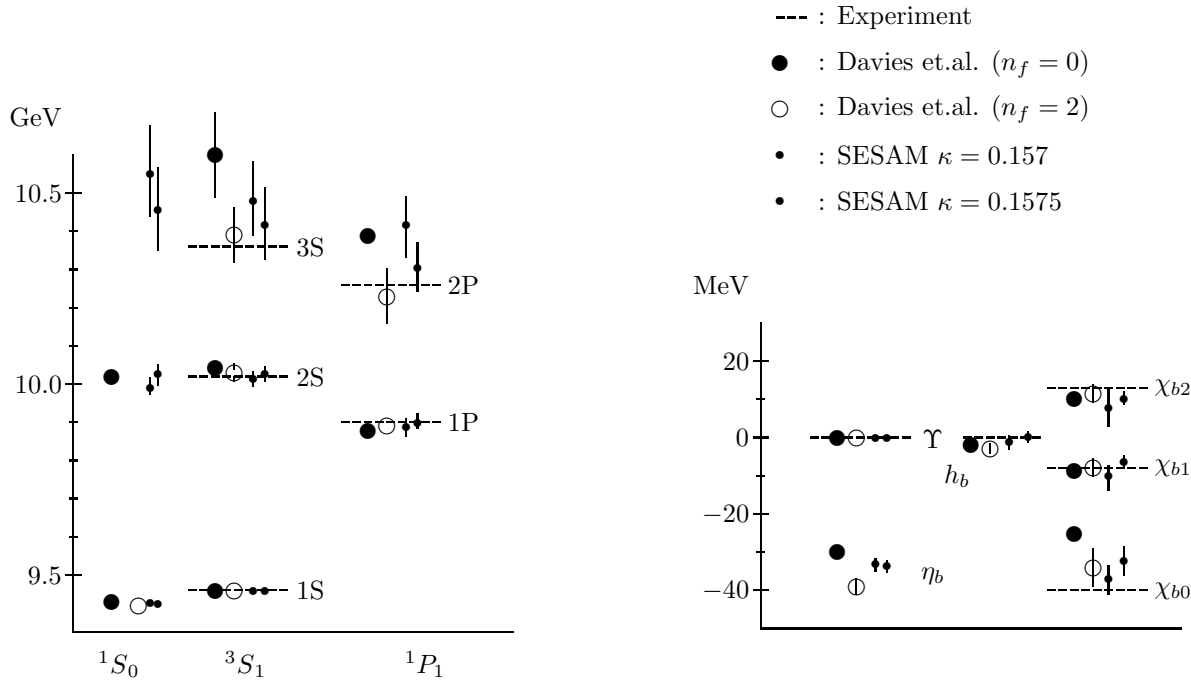

\setlength{\unitlength}{.02in}
\begin{picture}(125,140)(5,930)
\put(15,935){\line(0,1){125}}
\multiput(13,950)(0,50){3}{\line(1,0){4}}
\multiput(14,950)(0,10){10}{\line(1,0){2}}
\put(12,950){\makebox(0,0)[r]{9.5}}
\put(12,1000){\makebox(0,0)[r]{10.0}}
\put(12,1050){\makebox(0,0)[r]{10.5}}
\put(12,1070){\makebox(0,0)[r]{GeV}}
\put(15,935){\line(1,0){115}}

\put(27,930){\makebox(0,0)[t]{${^1S}_0$}}

\put(25,943.1){\circle*{4}}
\put(25,1002){\circle*{4}}
f
{\put(30,942){\circle{4}}}

{
\input{plot_11S0}
\input{plot_21S0}
\input{plot_31S0}
}

{
\input{plot_1575_21S0}
\input{plot_1575_11S0}
\input{plot_1575_31S0}
}


\put(58,930){\makebox(0,0)[t]{${^3S}_1$}}
\put(75,946){\makebox(0,0){1S}}
\multiput(43,946)(3,0){9}{\line(1,0){2}}
\put(75,1002){\makebox(0,0){2S}}
\multiput(43,1002)(3,0){9}{\line(1,0){2}}
\put(75,1036){\makebox(0,0){3S}}
\multiput(43,1036)(3,0){9}{\line(1,0){2}}

\put(50,946){\circle*{4}}
\put(50,1004.1){\circle*{4}}
\put(50,1060){\circle*{4}}
\put(50,1060){\line(0,1){11}}
\put(50,1060){\line(0,-1){11}}

{
\put(55,946){\circle{4}}
\put(55,1003){\circle{4}}
\put(55,1004){\line(0,1){1.4}}
\put(55,1002){\line(0,-1){1.4}}
\put(55,1039.1){\circle{4}}
\put(55,1039.1){\line(0,1){7.2}}
\put(55,1039.1){\line(0,-1){7.2}}
}

{
\put(60,946){\circle*{2}}
\input{plot_23S1}
\input{plot_33S1}
}

{
\put(63,946){\circle*{2}}
\input{plot_1575_23S1}
\input{plot_1575_33S1}
}


\put(105,930){\makebox(0,0)[t]{$^1P_1$}}

\put(115,990){\makebox(0,0){1P}}
\multiput(83,990)(3,0){9}{\line(1,0){2}}
\put(115,1026){\makebox(0,0){2P}}
\multiput(83,1026)(3,0){9}{\line(1,0){2}}

\put(90,987.6){\circle*{4}}
\put(90,1038.7){\circle*{4}}

{
\put(95,989){\circle{4}}
\put(95,1023){\circle{4}}
\put(95,1023){\line(0,1){7.2}}
\put(95,1023){\line(0,-1){7.2}}
}

{
\input{plot_11P1}
\input{plot_21P1}
}

{
\input{plot_1575_11P1}
\input{plot_1575_21P1}
}



\end{picture}
\hfill
\setlength{\unitlength}{.02in}
\begin{picture}(125,170)(15,-56)
\put(15,-50){\line(0,1){80}}
\multiput(13,-40)(0,20){4}{\line(1,0){4}}
\multiput(14,-40)(0,10){7}{\line(1,0){2}}
\put(12,-40){\makebox(0,0)[r]{$-40$}}
\put(12,-20){\makebox(0,0)[r]{$-20$}}
\put(12,0){\makebox(0,0)[r]{$0$}}
\put(12,20){\makebox(0,0)[r]{$20$}}
\put(12,40){\makebox(0,0)[r]{MeV}}
\put(15,-50){\line(1,0){110}}

\multiput(19,110)(3,0){3}{\line(1,0){2}}
\put(30,110){\makebox(0,0)[l]{: {Experiment}}}
\put(20, 98){\makebox(0,0)[tl]{\circle*{4}}}
\put(30, 98){\makebox(0,0)[l]{: {Davies et.al. $(n_f = 0)$}}}
{\put(20,86){\makebox(0,0)[tl]{\circle{4}}}}
\put(30,86){\makebox(0,0)[l]{: {Davies et.al. $(n_f = 2)$}}}
{\put(20,74){\makebox(0,0)[tl]{\circle*{2}}}}
\put(30,74){\makebox(0,0)[l]{: {SESAM $\kappa = 0.157$}}}
{\put(20,62){\makebox(0,0)[tl]{\circle*{2}}}}
\put(30,62){\makebox(0,0)[l]{: {SESAM $\kappa = 0.1575$}}}

\multiput(28,0)(3,0){9}{\line(1,0){2}}
\put(60,2){\makebox(0,0)[t]{$\Upsilon$}}
\put(60,-34){\makebox(0,0)[t]{$\eta_b$}}

\put(35,0){\circle*{4}}
\put(35,-29.9){\circle*{4}}

{
\put(40,0){\circle{4}}
\put(40,-39){\circle{4}}
\put(40,-39){\line(0,1){2}}
\put(40,-39){\line(0,-1){2}}
}
{
\put(45,0){\circle*{2}}
\input{plot_13S1-11S0}
}
{
\put(48,0){\circle*{2}}
\input{plot_1575_13S1-11S0}
}

\multiput(68,0)(3,0){7}{\line(1,0){2}}
\put(63,-10){\makebox(0,0)[l]{$h_b$}}
\put(70,-1.8){\circle*{4}}
{
\put(75,-2.9){\circle{4}}
\put(75,-2.9){\line(0,1){1.2}}
\put(75,-2.9){\line(0,-1){1.2}}
}

{\input{plot_1CG3P2-11P1}}
{\input{plot_1575_1CG3P2-11P1}}

\multiput(90,-40)(3,0){9}{\line(1,0){2}}
\put(118,-40){\makebox(0,0)[l]{$\chi_{b0}$}}
\multiput(90,-8)(3,0){9}{\line(1,0){2}}
\put(118,-8){\makebox(0,0)[l]{$\chi_{b1}$}}
\multiput(90,13)(3,0){9}{\line(1,0){2}}
\put(118,13){\makebox(0,0)[l]{$\chi_{b2}$}}

\put(97,-25.1){\circle*{4}}
\put(97,-8.6){\circle*{4}}
\put(97,10.2){\circle*{4}}

{
\put(102,-34){\circle{4}}
\put(102,-34){\line(0,1){5}}
\put(102,-34){\line(0,-1){5}}
\put(102,-7.9){\circle{4}}
\put(102,-7.9){\line(0,1){2.4}}
\put(102,-7.9){\line(0,-1){2.4}}
\put(102,11.5){\circle{4}}
\put(102,11.5){\line(0,1){2.4}}
\put(102,11.5){\line(0,-1){2.4}}
}

{
\input{plot_1CG3P2-13P0}
\input{plot_13P2-13P1}
\input{plot_13P2-13P0}
}

{
\input{plot_1575_1CG3P2-13P0}
\input{plot_1575_1CG3P2-13P1}
\input{plot_1575_13P2-1CG3P2}
}
\end{picture}
\caption{\small $\Upsilon$ Spectrum and Hyperfine Splittings:\label{SPLITT}
  Quenched: $\beta=6.0,16^3\times 32$ \cite{1}; Dynamical Staggered: $\beta
  = 5.6, am_q = 0.01, 16^3\times 32$ \cite{4}; Dynamical Wilson: $\beta =
  5.6, \kappa = 0.157, 0.1575, 16^3\times 32$. Errors are purely
  statistical. Note that Davies et al.\ have updated their quenched
  simulation meanwhile, see C. Davies, these proceedings.}
\end{figure*}

The lattice scale is taken from the average of the $2S-1S$ and
$CG(^3P)-1S$ splittings (see Tab.~9). We do not include ${\cal
  O}(a^2)$ gluonic corrections at this stage as their effect is not
visible with present statistic. For the ratio of splittings, $R =
(2^3S_1 - 1^3S1)/(CG(^3P) - 1^3S_1)$, we obtain a value $R = 1.28(10)$
at $\kappa = 0.1575$ and $R=1.30(9)$ at $\kappa=0.157$ which is
consistent with the experimental number $R = 1.28$. Tab.~10 lists our
results in lattice units, Figure \ref{SPLITT} sketches the $\Upsilon$
spectrum and hyperfine-splittings.
\begin{table}[htb]
\setlength{\tabcolsep}{.7pc}
\caption{Lattice Spacings.
\label{tab:space}}
\begin{tabular*}{.465\textwidth}{@{}|l@{\extracolsep{\fill}}|l|l|}
\hline
 {Lattice Spacings} & 
 $\scriptstyle \kappa = 0.157$ & $\scriptstyle \kappa = 0.1575$ \\
\hline
\hline
$\scriptstyle  a^{-1}\left( 2^3S_1 - 1^3S_1\right) [GeV] $ &
$ 2.34(15)$ & $ 2.48(14)$ \\
$\scriptstyle a^{-1}\left( 1^3S_1 - 1 CG(^3P)\right) [GeV] $ & 
$2.40(13)$ & $ 2.50(11)$ \\
\hline
\end{tabular*}
\end{table}
\begin{table}[htb]
\setlength{\tabcolsep}{1pc}
\caption{Fit results for radial and spin splittings.
\label{tab:res}}
\begin{tabular*}{.465\textwidth}{@{}|l@{\extracolsep{\fill}}|l|l|}
\hline
$ aM_b^0 = 1.7 $ & $\kappa=0.157$  & $\kappa=0.1575$ \\
\hline\hline
$ 2\; ^1S_0 - 1\; ^1S_0 $ & 0.245(8)  & 0.240(7) \\ 
\hline
$ 3\; ^1S_0 - 1\; ^1S_0 $ & 0.46(5)    & 0.41(4)  \\ 
\hline
$ 2\; ^3S_1 - 1\; ^3S_1 $ & 0.241(13)  & 0.226(15) \\ 
\hline
$ 3\; ^3S_1 - 1\; ^3S_1 $ & 0.41(2)    & 0.38(3)   \\ 
\hline
$ 1\; ^1P_1 - 1\; ^3S_1 $ & 0.186(8)   & 0.176(9) \\
\hline
$ 2\; ^1P_1 - 1\; ^1P_1 $ & 0.23(2)    & 0.18(3)  \\
\hline
$ 1\; ^3S_1 - 1\; ^1S_0 $ & 0.0142(2)  & 0.0135(2) \\
\hline
$ 1\; ^3P_2 - 1\; ^1P_1 $ & 0.0032(11)  & 0.0039(5) \\
\hline
$ 1\; ^1P_1 - 1\; ^3P_1 $ & 0.0040(10)  & 0.0026(7) \\
\hline
$ 1\; ^1P_1 - 1\; ^3P_0 $ & 0.015(4)   & 0.013(2) \\
\hline
\end{tabular*}
\end{table}

\subsection{Remarks}

We have presented preliminary results on the bottomonium spectrum
calculated from NRQCD, in a gauge field background with 2 flavours of
dynamical Wilson quarks. Radial excitations and hyperfine splittings
have been determined for two ensembles of configurations corresponding
to two different sea-quark masses. We observe better agreement with
experiment as compared to the quenched result confirming the results
of simulations with staggered fermions. Up to now we are, however, not
able to give a conclusive statement concerning the dependence of
energy levels on the dynamical quark mass. To decide on this issue and
to disentangle unquenching effects from relativistic corrections we
are going ({\em i}) to complete the analysis with a third
$\kappa$-value ($\kappa_{sea} = 0.156$), ({\em ii}) we will carry out
a quenched simulation for our choice of heavy quark action and ({\em
  iii}) we will exploit T$\chi$L configurations at
$\kappa_{sea}=0.158$.

\section*{SUMMARY}

I have tried to give an impression of the goals and achievements of
the \sesam\ and the \tkl\ projects, simulations with Wilson fermions
at intermediate lattice sizes and dynamical quark masses. To set a
significant landmark we have concentrated on one $\beta$ value.

It would of course be highly desirable to carry out a scaling analysis
in full QCD, {\em i.e.}, the extrapolation to the continuum limit.
Such a study, making use of improved actions, will be the adequate
computational challenge for computers of $\ge$ CP-PACS performance!

\vspace{12pt}
\noindent {\bf Acknowledgments.}
We are indepted to Prof.\ A.\ Frommer and his group for a very pleasant and
fruitful cooperation.  Th.\ L.\ and K.\ S.\ thank Profs.\ Y.\ Iwasaki and A.\
Ukawa for their hospitality and a beautiful workshop at Tsukuba Center
for Computational Physics.  The work of Th.\ L.\ is supported by
the Deutsche Forschungsgemeinschaft DFG under grant No.\ Schi 257/5-1.

\end{document}